*Florentin Smarandache   Sukanto Bhattacharya*

*Mohammad Khoshnevisan*

editors

**Computational Modeling in Applied Problems:**

**collected papers on econometrics, operations research,**

**game theory and simulation**

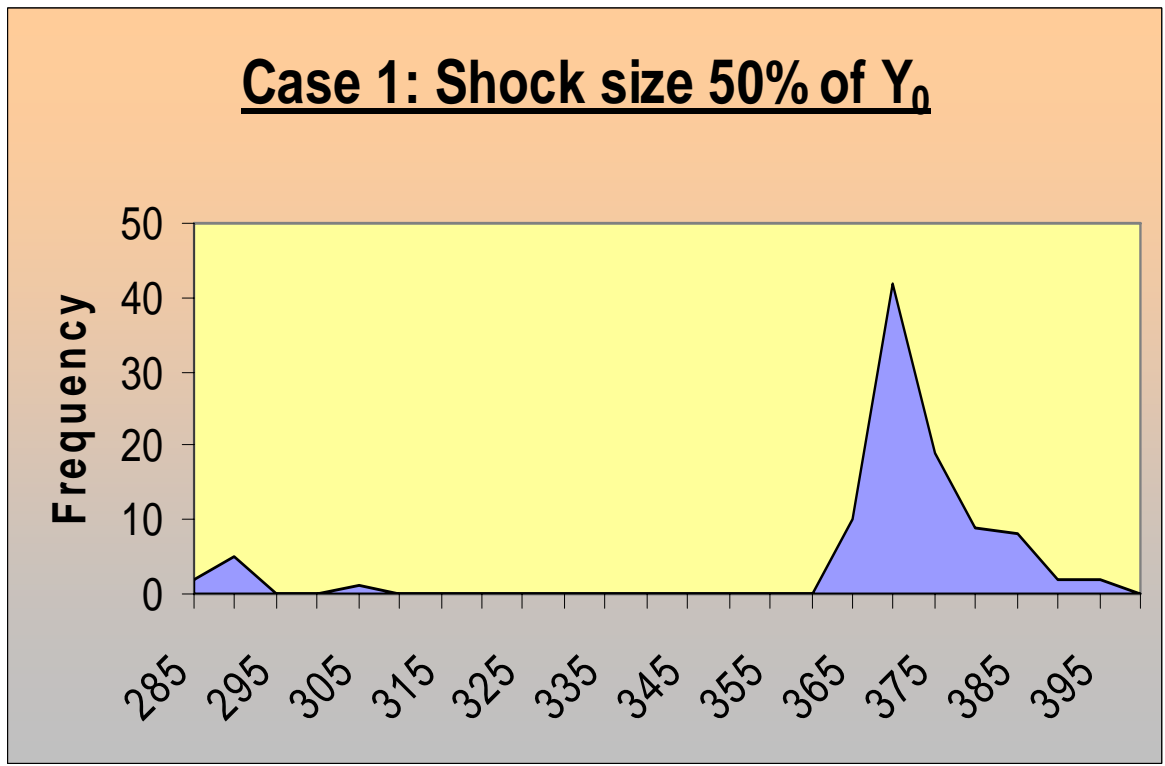

**2006**

*F*LORENTIN *S*MARANDACHE    *S*UKANTO *B*HATTACHARYA

*M*OHAMMAD *K*HOSHNEVISAN

editors

**2006**

**Contents**

Forward ….. 3

**Forward**

Computational models pervade all branches of the exact sciences and have in recent times also started to prove to be of immense utility in some of the traditionally 'soft' sciences like ecology, sociology and politics. This volume is a collection of a few cutting-edge research papers on the application of variety of computational models and tools in the analysis, interpretation and solution of vexing real-world problems and issues in economics, management, ecology and global politics by some prolific researchers in the field.

The Editors

# Econometric Analysis on Efficiency of Estimator

M. Khoshnevisan

Griffith University, School of Accounting and Finance, Australia

F. Kaymram

Massachusetts Institute of Technology

Department of Mechanical Engineering, USA

{currently at Sharif University, Iran}

Housila P. Singh, Rajesh Singh

Vikram University, Department of Mathematics and Statistics, India

F. Smarandache

Department of Mathematics, University of New Mexico, Gallup, USA

**Abstract**

This paper investigates the efficiency of an alternative to ratio estimator under the super population model with uncorrelated errors and a gamma-distributed auxiliary variable. Comparisons with usual ratio and unbiased estimators are also made.

## 1. Introduction

It is well known that the ratio method of estimation occupies an important place in sample surveys. When the study variate y and the auxiliary variate x is positively (high) correlated, the ratio method of estimation is quite effective in estimating the population mean of the study variate y utilizing the information on auxiliary variate x.

Consider a finite population with N units and let $x_i$ and $y_i$ denote the values for two positively correlated variates x and y respectively for the ith unit in this population, i=1,2,…,N. Assume that the population mean $\bar{X}$ of x is known. Let $\bar{x}$ and $\bar{y}$ be the sample means of x and y respectively based on a simple random sample of size n (n < N) units drawn without replacement scheme. Then the classical ratio estimator for $\bar{Y}$ is defined by

$$\bar{y}_r = \bar{y}(\bar{X}/\bar{x}) \tag{1.1}$$

The bias and mean square error (MSE) of $\bar{y}_r$ are, up to second order moments,

$$B(\bar{y}_r) = \lambda(R S^2_x - S_{yx})/\bar{X} \tag{1.2}$$

$$M(\bar{y}_r) = \lambda(S^2_y + R^2 S^2_x - 2R S_{yx}), \tag{1.3}$$

where $\lambda = (N-n)/(nN)$,

$R = \bar{Y}/\bar{X}$, $S^2_y = (N-1)^{-1}\sum_{i=1}^{N}(y_i - \bar{Y})^2$, $s^2_x = (N-1)^{-1}\sum_{i=1}^{N}(x_i - \bar{X})^2$,

and $S_{yx} = (N-1)^{-1}\sum_{i=1}^{N}(y_i - \bar{Y})(x_i - \bar{X})$.

It is clear from (1.3) that $M(\bar{y}_r)$ will be minimum when

$$R = S_{yx}/S^2_x = \beta, \tag{1.4}$$

where $\beta$ is the regression coefficient of y on x. Also for $R = \beta$,

the bias of $\bar{y}_r$ in (1.2) is zero. That is, $\bar{y}_r$ is almost unbiased for $\bar{Y}$.

Let $E(\bar{y}|\bar{x}) = \alpha + \beta \bar{x}$ be the line of regression of $\bar{y}$ on $\bar{x}$, where E denotes averaging over all possible sample design simple random sampling without replacement (SRSWOR). Then $\beta = S_{yx}/S_x^2$ and $\bar{Y} = \alpha + \beta \bar{X}$ so that, in general,

$$R = (\alpha/\bar{X}) + \beta \tag{1.5}$$

It is obvious from (1.4) and (1.5) that any transformation that brings the ratio of population means closer to $\beta$ will be helpful in reducing the mean square error (MSE) as well as the bias of the ratio estimator $\bar{y}_r$. This led Srivenkataramana and Tracy (1986) to suggest an alternative to ratio estimator $\bar{y}_r$ as

$$\bar{y}_a = \bar{z}(\bar{X}/\bar{x}) + A = \bar{y}_r - A\{(\bar{X}/\bar{x}) - 1\} \tag{1.6}$$

which is based on the transformation

$$\bar{z} = \bar{y} - A, \tag{1.7}$$

where $E(\bar{z}) = \bar{Z}(= \bar{Y} - A)$ and A is a suitably chosen scalar.

In this paper exact expressions of bias and MSE of $\bar{y}_a$ are worked out under a super population model and compared with the usual ratio estimator.

## 2. The Super Population Model

Following Durbin (1959) and Rao (1968) it is assumed that the finite population under consideration is itself a random sample from a super population and the relation between x and y is of the form:

$$y_i = \alpha + \beta x_i + u_i \; ; \quad (i = 1,2,\ldots,N) \tag{1.8}$$

where $\alpha$ and $\beta$ are unknown real constants; $u_i$'s are uncorrelated random errors with conditional (given $x_i$) expectations

$$E(u_i|x_i) = 0 \tag{1.9}$$

$$E(u_i^2|x_i) = \delta \, x_i^g \tag{1.10}$$

( $i=1,2,\ldots,N$), $o \langle \delta \langle \infty$, $o \leq g \leq 2$ and $x_i$ are independently identically

distributed ( i.i.d.) with a common gamma density

$$G(\theta) = e^{-x} x^{\theta-1} / \Gamma\theta, \; x \rangle o, \; 2 \langle \theta \langle \infty . \tag{2.1}$$

We will write $E_x$ to denote expectation operator with respect to the common distribution of $x_i$ (i=1,2,3,…,N) and $E_x E_c$, as the over all expectation operator for the model. We denote a design by p and the design expectation $E_p$, for instance, see Chaudhuri and Adhikary (1983,89) and Shah and Gupta (1987). Let 's' denote a simple random sample of N distinict labels chosen without replacement out of i=1,2,3……N. Then

$$X(=N \bar{X}) = \sum_{i \in s} x_i + \sum_{i \notin s} x_i \tag{2.2}$$

Following Rao and Webster (1966) we will utilize the distributional properties of $x_j / x_i$, $\sum_{i \in s} x_i$, $\sum_{i \notin s} x_i$, $\sum_{i \in s} x_i / \sum_{i \notin s} x_i$ in our subsequent derivations.

## 3. The bias and mean square error

The estimator $\bar{y}_a$ in (1.6) can be written as

$$\bar{y}_a = \left[ (1/n)\left(\sum_{i \in s} y_i\right) \frac{\left(n\sum_{i=1}^{N} x_i\right)}{\left(N\sum_{i \in s} x_i\right)} - A\left\{\frac{\left(n\sum_{i=1}^{N} x_i\right)}{\left(N\sum_{i \in s} x_i\right)} - 1\right\} \right] \quad (3.1)$$

based on a simple random sample of n distinct labels chosen without replacement out of i = 1,2,…,N.

The bias

$$B = E_p ( \bar{y}_a - \bar{Y} ) \quad (3.2)$$

of $\bar{y}_a$ has model expectation $E_m(B)$ which works out as follows:

$$E_m ( B ( \bar{y}_a )) = E_p E_x E_c \left[ \left\{\alpha + \beta(1/n)\left(\sum_{i \in s} x_i\right) + \bar{u}\right\} \frac{n\sum_{i=1}^{N} x_i}{n\sum_{i \in s} x_i} \right.$$

$$- A \left\{\frac{n\left(\sum_{i=1}^{N} x_i\right) - 1}{N\left(\sum_{i \in s} x_i\right)}\right\} -$$

$$- E_x E_c ( \alpha + \beta \bar{x} + \bar{U} )$$

$$= E_p E_x E_c$$

$$\left[ \alpha\left(n\sum_{i=1}^{N} x_i / N\sum_{i \in s} x_i\right) + \beta(1/N)\left(\sum_{i=1}^{N} x_i\right) + \left(\sum_{i \in s} u_i\right)\left(\sum_{i=1}^{N} x_i / N\sum_{i \in s} x_i\right) - A\left\{\left(n\sum_{i=1}^{N} x_i\right) / \left(N\sum_{i \in s} x_i\right) - 1\right\} \right]$$

$$- \text{E}_x \text{E}_c\ (\alpha + \beta \bar{X})$$

$$= \text{E}_p \text{E}_x \left[ \alpha \left( n \sum_{i=1}^{N} x_i \right) \bigg/ \left( N \sum_{i \in s} x_i \right) + \beta \bar{X} - A \left\{ \left( n \sum_{i=1}^{N} x_i \right) \bigg/ \left( N \sum_{i \in s} x_i \right) - 1 \right\} \right] - \alpha - \beta\ \text{E}_x\ (\bar{X})$$

$$= \text{E}_x \left[ \alpha (n/N) \left( 1 + \sum_{i \notin s} x_i \bigg/ \sum_{i \in s} x_i \right) - A \left\{ (n/N) \left( 1 + \sum_{i \notin s} x_i \bigg/ \sum_{i \in s} x_i \right) - 1 \right\} \right] - \alpha$$

$$= \alpha (n/N)\{1 + (N-n)\theta/(n\theta - 1)\}$$
$$\quad -A\ \{(n/N)(1 + (N-n)\theta/(n\theta - 1)) - 1\} - \alpha$$

$$= \alpha [(n/N - 1) + \{n(N-n)\theta/N(n\theta - 1)\}]$$
$$\quad -A\ [-(N-n)/N + \{(N-n)n\theta/N(n\theta - 1)\}]$$

$$= (N-n)\ (\alpha - A)/N(n\theta - 1) \qquad (3.3)$$

For SRSWOR sampling scheme, the mean square error

$$\text{M}\ (\bar{y}_a) = \text{E}_p\ (\bar{y}_a - \bar{Y})^2 \qquad (3.4)$$

of $\bar{y}_a$ has the following formula for model expectations

$\text{E}_m\ (\ \text{M}\ (\bar{y}_a))\ :$

$$\text{E}_m(M(\bar{y}_a)) = \left[ \text{E}_m(M(\bar{y}_r)) + (N-n)(Nn\theta + 2N - 2n)(A^2 - 2A\alpha)/N^2(n\theta - 1)(n\theta - 2) \right]$$
$$(3.5)$$

where

$$M(\bar{y}_r) = E_p(\bar{y}_r - \bar{Y})^2 \tag{3.6}$$

is the MSE of $\bar{y}_r$ under SRSWOR scheme has the model expectation

$$E_m(M(\bar{y}_r)) = \{(N-n)/N^2\}$$
$$\left[ \left\{ \frac{(Nn\theta + 2N - 2n)\alpha^2}{(n\theta - 1)(n\theta - 2)} \right\} + \frac{\delta\left\{(n\theta + g - 1)(n\theta + g - 2) + n\theta(N\theta - n\theta + 1)\right\}}{(n\theta + g - 1)(n\theta + g - 2)} \frac{\Gamma(\theta + g)}{\Gamma\theta} \right] \tag{3.7}$$

$[See, Rao(1968, p.439)]$

Further, we note that for SRSWOR sampling scheme, the bias

$$B(\bar{y}_r) = E_p(\bar{y}_r - \bar{Y}) \tag{3.8}$$

of usual ratio estimator has the model expectation

$$E_m(B(\bar{y}_r)) = (N-n)\alpha / (n\theta - 1) \tag{3.9}$$

*We note from (3.3) and (3.9) that*

$$|E_m(B(\bar{y}_a))| \langle |E_m(B(\bar{y}_r))|$$

if
$$|(\alpha - A)| \langle |\alpha|$$

or if
$$(\alpha - A)^2 \langle \alpha^2$$

or if
$$o \langle A \langle 2\alpha \tag{3.10}$$

*Further we have from (3.5) that*

$$E_m(M(\bar{y}_a)) - E_m(M(\bar{y}_r)) < o$$

*if*

$$(A^2 - 2A\alpha) < o$$

or if

$$o \langle A \langle 2\alpha \qquad (3.11)$$

which is the same as in (3.10).

Thus we state the following theorem:

Theorem 3.1 : The estimator $\bar{y}_a$ is less biased as well as more efficient than usual ratio estimator $\bar{y}_r$ if

$$o \langle A \langle 2\alpha \qquad (\alpha \neq o)$$

i . e . when A lies between $o$ and $2\alpha$.

Therefore, when intercept term $\alpha(\neq o)$ in the model (2.1) is sizable, there will be sufficient flexibility in picking A.

It is to be noted that for $\alpha = o, \bar{y}_r$ is unbiased and efficient than $\bar{y}_a$.

The minimization of (3.5) with respect to A leads to

$$A = \alpha = A_{opt} \text{ (say)} \qquad (3.12)$$

Substitution of (3.12) in (3.5) yields the minimum value of

$E_m(M(\bar{y}_a))$ as

$$\text{min. } E_m(M(\bar{y}_a)) = \frac{(N-1)}{N^2} \frac{\delta[(n\theta + g - 1)(n\theta + g - 2) + n\theta(N\theta - n\theta + 1)]}{(n\theta + g - 1)(n\theta + g - 2)} \frac{\Gamma(\theta + g)}{\Gamma\theta}$$

(3.13)

which equals to $E_m(M(\bar{y}_r))$ when $\alpha = o$.

It is interesting to note that when $A = \alpha$, $\bar{y}_a$ is unbiased and attained its minimum average MSE in model (2.1).

In practice the value of $\alpha$ will have to be assessed, at the estimation stage, to be used as A. To assess $\alpha$, we may use scatter diagram of y versus x for data from a pilot study, or a part of the data from the actual study and judge the y-intercept of the best fitting line.

From (3.7) and (3.13) we have

$$E_m(M(\bar{y}_r)) - \min. E_m(M(\bar{y}_a)) = \{(N-n)(Nn\theta + 2N - 2n)\alpha^2\}/\{N^2(n\theta-1)(n\theta-2)\} > o$$

(3.14)

which shows that $\bar{y}_a$ is more efficient than ratio estimator when $A = \alpha$

is known exactly. For $\alpha = o$

$$\min. E_m(M(\bar{y}_a)) = E_m(M(\bar{y}_r)) \qquad (3.15)$$

For SRSWOR, the variance

$$V(\bar{y}) = E_p(\bar{y} - \bar{Y})^2 \qquad (3.16)$$

of usual unbiased estimator has the model expectation:

$$E_m(V(\bar{y})) = (N-n)[\beta^2\theta + \{\delta\Gamma(\theta+g)/\Gamma\theta\}]/nN \qquad (3.17)$$

The expressions of $E_m(M(\bar{y}_a))$ and $E_m(V(\bar{y}))$ are not easy task to compare algebraically. Therefore in order to facilitate the comparison, denoting

$$E_1 = 100 E_m(V(\bar{y}))/E_m(M(\bar{y}_a)) \text{ and } E_2 = 100 E_m(V(\bar{y}_r))/E_m(M(\bar{y}_a)),$$

we present below in tables 1,2,3, the values of the relative efficiencies of $\bar{y}_a$ with respect to $\bar{y}$ and $\bar{y}_r$ for a few combination of the parametric values under the model (2.1). Values are given for N = 60, $\delta = 2.0, \theta = 8, \alpha = 0.5$, 1.0, 1.5, $\beta = 0.5, 1.0, 1.5$ and g = 0.0, 0.5, 1.0, 1.5, 2.0. The ranges of A, for $\bar{y}_a$ to be better than $\bar{y}_r$ for given $\alpha = 0.5, 1.0, 1.5$ are respectively (0,1), (0,2), (0,3). This clearly indicates that as the size of

$\alpha$ increases the range of A for $\bar{y}_a$ to be better than $\bar{y}_r$ increases i.e. flexibility of choosing A increases.

We have made the following observations from the tables 1,2 and 3 :

(i) As g increases both $E_1$ and $E_2$ decrease. When n increases $E_1$ increases while $E_2$ decreases.

(ii) As $\alpha$ increases ( i.e. if the intercept term $\alpha$ departs from origin in positive direction) relative efficiency of $\bar{y}_a$ with respect to $\bar{y}$ decreases while $E_2$ increases.

(iii) As $\beta$ increases $E_1$ increases for fixed g while $E_2$ is unaffected.

(iv) The maximum gain in efficiency is observed over $\bar{y}$ as well as over $\bar{y}_r$ if A coincide with the value of $\alpha$. Finally, the estimator $\bar{y}_a$ is to be preferred when the intercept term $\alpha$ departs substantially from origin.

**Table 1: Relative efficiencies of $\bar{y}_a$ with respect to $\bar{y}$ and $\bar{y}_\Gamma$**

$\alpha = 0.5$

| g | $\beta$ | n = 10 | | | | | |
|---|---|---|---|---|---|---|---|
| | | $E_1$ | | | $E_2$ | | |
| | | A | | | A | | |
| | | 0.30 | 0.60 | 0.90 | 0.30 | 0.60 | 0.90 |
| 0.0 | 0.5 | 192.86 | 193.23 | 191.40 | 101.34 | 101.54 | 100.57 |
| | 1.0 | 482.16 | 483.16 | 478.09 | 101.34 | 101.54 | 100.57 |
| | 1.5 | 964.32 | 966.17 | 956.98 | 101.34 | 101.54 | 100.57 |
| 0.5 | 0.5 | 132.67 | 132.77 | 132.30 | 100.49 | 100.56 | 100.21 |
| | 1.0 | 237.82 | 237.99 | 237.16 | 100.49 | 100.56 | 100.21 |
| | 1.5 | 413.08 | 413.36 | 411.93 | 100.49 | 100.56 | 100.21 |
| 1.0 | 0.5 | 111.06 | 111.08 | 110.95 | 10.17 | 100.19 | 100.07 |
| | 1.0 | 148.08 | 148.11 | 147.93 | 10.17 | 100.19 | 100.07 |
| | 1.5 | 209.78 | 209.83 | 209.57 | 10.17 | 100.19 | 100.07 |
| 1.5 | 0.5 | 103.99 | 104.00 | 103.96 | 100.06 | 100.07 | 100.03 |
| | 1.0 | 116.64 | 116.65 | 116.60 | 100.06 | 100.07 | 100.03 |
| | 1.5 | 137.71 | 137.72 | 137.66 | 100.06 | 100.07 | 100.03 |
| 2.0 | 0.5 | 102.23 | 102.23 | 102.22 | 100.02 | 100.02 | 100.01 |
| | 1.0 | 106.43 | 106.43 | 106.42 | 100.02 | 100.02 | 100.01 |
| | 1.5 | 113.43 | 113.43 | 113.42 | 100.02 | 100.02 | 100.01 |

| g | $\beta$ | n = 20 | | | | | |
|---|---|---|---|---|---|---|---|
| | | $\alpha = 0.5$ | | | | | |
| | | $E_1$ | | | $E_2$ | | |
| | | A | | | A | | |
| | | 0.30 | 0.60 | 0.90 | 0.30 | 0.60 | 0.90 |
| 0.0 | 0.5 | 196.58 | 196.96 | 195.11 | 103.33 | 101.52 | 100.56 |
| | 1.0 | 491.46 | 492.39 | 487.77 | 103.33 | 101.52 | 100.56 |
| | 1.5 | 982.92 | 984.39 | 975.53 | 103.33 | 101.52 | 100.56 |
| 0.5 | 0.5 | 134.37 | 134.46 | 134.46 | 100.48 | 100.55 | 100.20 |
| | 1.0 | 240.86 | 241.02 | 240.02 | 100.48 | 100.55 | 100.20 |
| | 1.5 | 418.35 | 418.63 | 417.20 | 100.48 | 100.55 | 100.20 |
| 1.0 | 0.5 | 111.76 | 111.79 | 111.65 | 100.17 | 100.19 | 100.07 |
| | 1.0 | 149.01 | 149.05 | 148.87 | 100.17 | 100.19 | 100.07 |
| | 1.5 | 211.10 | 211.16 | 210.90 | 100.17 | 100.19 | 100.07 |
| 1.5 | 0.5 | 104.00 | 104.00 | 103.96 | 100.06 | 100.07 | 100.02 |
| | 1.0 | 116.64 | 116.65 | 116.60 | 100.06 | 100.07 | 100.02 |
| | 1.5 | 137.71 | 137.73 | 137.67 | 100.06 | 100.07 | 100.02 |
| 2.0 | 0.5 | 101.60 | 101.60 | 101.58 | 100.02 | 100.02 | 100.01 |
| | 1.0 | 105.77 | 105.77 | 105.76 | 100.02 | 100.02 | 100.01 |
| | 1.5 | 112.73 | 112.73 | 112.73 | 100.02 | 100.02 | 100.01 |

**Table 2: Relative efficiencies of $\bar{y}_a$ with respect to $\bar{y}$ and $\bar{y}_r$**

| $\alpha = 1.0$ | | | | | | | | | |
|---|---|---|---|---|---|---|---|---|---|
| g | $\beta$ | \multicolumn{8}{c}{n = 10} |
| | | $E_1$ A | | | | $E_2$ A | | | |
| | | 0.50 | 1.0 | 1.50 | 1.90 | 0.50 | 1.0 | 1.50 | 1.90 |
| 0.0 | 0.5 | 190.31 | 193.36 | 190.31 | 183.82 | 104.73 | 106.41 | 104.73 | 101.16 |
| | 1.0 | 475.78 | 483.40 | 475.78 | 459.55 | 104.73 | 106.41 | 104.73 | 101.16 |
| | 1.5 | 951.55 | 966.79 | 951.55 | 919.10 | 104.73 | 106.41 | 104.73 | 101.16 |
| 0.5 | 0.5 | 132.03 | 132.80 | 132.03 | 130.34 | 101.73 | 102.32 | 101.73 | 100.43 |
| | 1.0 | 236.67 | 238.05 | 236.67 | 233.65 | 101.73 | 102.32 | 101.73 | 100.43 |
| | 1.5 | 411.07 | 413.46 | 411.07 | 405.82 | 101.73 | 102.32 | 101.73 | 100.43 |
| 1.0 | 0.5 | 110.87 | 111.09 | 110.87 | 110.36 | 100.61 | 100.82 | 100.61 | 100.15 |
| | 1.0 | 147.82 | 148.12 | 147.82 | 147.15 | 100.61 | 100.82 | 100.61 | 100.15 |
| | 1.5 | 209.42 | 209.84 | 209.42 | 208.46 | 100.61 | 100.82 | 100.61 | 100.15 |
| 1.5 | 0.5 | 103.93 | 104.00 | 103.93 | 103.77 | 100.21 | 100.28 | 100.21 | 100.05 |
| | 1.0 | 116.57 | 116.65 | 116.57 | 116.39 | 100.21 | 100.28 | 100.21 | 100.05 |
| | 1.5 | 137.63 | 137.73 | 137.63 | 137.41 | 100.21 | 100.28 | 100.21 | 100.05 |
| 2.0 | 0.5 | 102.21 | 102.23 | 102.21 | 102.15 | 100.67 | 100.09 | 100.07 | 100.01 |
| | 1.0 | 106.41 | 106.43 | 106.41 | 106.3 | 100.67 | 100.09 | 100.07 | 100.01 |
| | 1.5 | 113.41 | 113.43 | 113.41 | 113.35 | 100.67 | 100.09 | 100.07 | 100.01 |

| g | $\beta$ | \multicolumn{8}{c|}{n = 20} |
|---|---|---|---|---|---|---|---|---|---|
| | | $E_1$ A | | | | $E_2$ A | | | |
| | | 0.50 | 1.0 | 1.50 | 1.90 | 0.50 | 1.0 | 1.50 | 1.90 |
| 0.0 | 0.5 | 194.01 | 197.08 | 194.01 | 187.47 | 104.67 | 106.33 | 104.67 | 101.14 |
| | 1.0 | 485.03 | 492.70 | 485.03 | 468.68 | 104.67 | 106.33 | 104.67 | 101.14 |
| | 1.5 | 970.06 | 985.40 | 970.06 | 937.36 | 104.67 | 106.33 | 104.67 | 101.14 |
| 0.5 | 0.5 | 133.73 | 134.49 | 133.73 | 132.05 | 101.70 | 102.28 | 101.70 | 100.08 |
| | 1.0 | 239.71 | 241.08 | 239.71 | 236.71 | 101.70 | 102.28 | 101.70 | 100.08 |
| | 1.5 | 416.35 | 418.73 | 416.35 | 411.13 | 101.70 | 102.28 | 101.70 | 100.08 |
| 1.0 | 0.5 | 111.07 | 111.08 | 111.07 | 111.08 | 100.60 | 100.80 | 100.60 | 100.15 |
| | 1.0 | 148.77 | 149.06 | 148.77 | 148.11 | 100.60 | 100.80 | 100.60 | 100.15 |
| | 1.5 | 210.75 | 211.17 | 210.75 | 209.82 | 100.60 | 100.80 | 100.60 | 100.15 |
| 1.5 | 0.5 | 103.94 | 104.01 | 103.94 | 103.78 | 100.20 | 100.27 | 100.20 | 100.05 |
| | 1.0 | 116.57 | 116.65 | 116.57 | 116.40 | 100.20 | 100.27 | 100.20 | 100.05 |
| | 1.5 | 137.64 | 137.73 | 137.64 | 137.42 | 100.20 | 100.27 | 100.20 | 100.05 |
| 2.0 | 0.5 | 101.58 | 101.60 | 101.58 | 101.52 | 100.07 | 100.09 | 100.07 | 100.01 |
| | 1.0 | 105.75 | 105.77 | 105.75 | 105.70 | 100.07 | 100.09 | 100.07 | 100.01 |
| | 1.5 | 112.71 | 112.73 | 112.71 | 112.65 | 100.07 | 100.09 | 100.07 | 100.01 |

$\alpha = 1.0$

**Table 3: Relative efficiencies of $\bar{y}_a$ with respect to $\bar{y}$ and $\bar{y}_r$**

| $\alpha = 1.5$ | | | | | | | | | | | |
|---|---|---|---|---|---|---|---|---|---|---|---|
| g | $\beta$ | \multicolumn{10}{c}{n = 10} |
| | | E₁ A | | | | | E₂ A | | | | |
| | | 0.60 | 1.20 | 1.80 | 2.40 | 2.90 | 0.60 | 1.20 | 1.80 | 2.40 | 2.90 |
| 0.0 | 0.5 | 183.82 | 192.25 | 192.25 / 171.79 | | 183.82 | 108.77 | 113.76 | 113.76 | 108.77 / 101.65 | |
| | 1.0 | 459.55 | 480.62 | 480.62 / 429.47 | | 459.55 | 108.77 | 113.76 | 113.76 | 108.77 / 101.65 | |
| | 1.5 | 919.10 | 961.25 | 961.25 / 858.94 | | 919.10 | 108.77 | 113.76 | 113.76 | 108.77 / 101.65 | |
| 0.5 | 0.5 | 130.34 | 132.52 | 132.52 / 127.01 | | 130.34 | 103.29 | 105.01 | 105.01 | 103.29 / 100.64 | |
| | 1.0 | 233.64 | 237.55 | 237.55 / 227.67 | | 233.65 | 103.29 | 105.01 | 105.01 | 103.29 / 100.64 | |
| | 1.5 | 405.82 | 412.60 | 412.60 / 395.44 | | 405.82 | 103.29 | 105.01 | 105.01 | 103.29 / 100.64 | |
| 1.0 | 0.5 | 110.36 | 111.01 | 111.01 / 109.34 | | 110.36 | 101.17 | 101.77 | 101.77 | 101.17 / 100.23 | |
| | 1.0 | 147.15 | 148.02 | 148.02 / 147.79 | | 147.15 | 101.17 | 101.77 | 101.77 | 101.17 / 100.23 | |
| | 1.5 | 208.46 | 209.69 | 209.69 / 206.53 | | 208.46 | 101.17 | 101.77 | 101.77 | 101.17 / 100.23 | |
| 1.5 | 0.5 | 103.77 | 103.98 | 103.98 / 103.44 | | 103.77 | 100.40 | 100.60 | 100.60 | 100.40 / 100.08 | |
| | 1.0 | 116.39 | 116.62 | 116.62 | | 116.39 | 100.40 | 100.60 | 100.60 | 100.40 | |

| | 1.5 | 137.41 | 137.69 | 116.01<br>137.69<br>139.68 | | 137.41 | 100.40 | 100.60 | 100.60 | 100.08<br>100.40<br>100.08 |
| --- | --- | --- | --- | --- | --- | --- | --- | --- | --- | --- |
| | 0.5 | 102.15 | 102.22 | 102.22<br>102.04 | | 102.15 | 100.13 | 100.20 | 100.20 | 100.13<br>100.03 |
| 2.0 | 1.0 | 106.35 | 106.42 | 106.42<br>106.24 | | 106.35 | 100.13 | 100.20 | 100.20 | 100.13<br>100.03 |
| | 1.5 | 113.35 | 113.42 | 113.42<br>113.23 | | 113.35 | 100.13 | 100.20 | 100.20 | 100.13<br>100.03 |

| $\alpha = 1.5$ | | | | | | | | | | | |
| --- | --- | --- | --- | --- | --- | --- | --- | --- | --- | --- | --- |
| G | $\beta$ | \multicolumn{10}{c}{n = 20} |
| | | $E_1$<br>A | | | | | $E_2$ | | | | |
| | | 0.60 | 1.20 | 1.80 | 2.40 | 2.90 | 0.60 | 1.20 | 1.80 | 2.40 | 2.90 |
| 0.0 | 0.5 | 187.47 | 196.97 | 195.97<br>175.33 | | 187.47 | 108.67 | 113.59 | 113.59 | 108.67<br>101.63 | |
| | 1.0 | 468.68 | 489.91 | 489.91<br>438.34 | | 468.68 | 108.67 | 113.59 | 113.59 | 108.67<br>101.63 | |
| | 1.5 | 937.36 | 979.83 | 979.83<br>876.67 | | 937.36 | 108.67 | 113.59 | 113.59 | 108.67<br>101.63 | |
| 0.5 | 0.5 | 132.05 | 134.21 | 134.21<br>128.73 | | 132.05 | 103.23 | 104.92 | 104.92 | 103.23<br>100.63 | |
| | 1.0 | 236.70 | 240.58 | 240.58<br>230.76 | | 236.70 | 103.23 | 104.92 | 104.92 | 103.23<br>100.63 | |
| | 1.5 | 411.13 | 417.87 | 417.87<br>400.80 | | 411.13 | 103.23 | 104.92 | 104.92 | 103.23<br>100.63 | |

| | | | | | | | | | |
|---|---|---|---|---|---|---|---|---|---|
| 1.0 | 0.5 | 111.08 | 111.72 | 111.72 110.08 | 111.08 | 101.14 | 101.72 | 101.72 | 101.14 100.23 |
| | 1.0 | 148.11 | 148.96 | 148.96 146.77 | 148.11 | 101.14 | 101.72 | 101.72 | 101.14 100.23 |
| | 1.5 | 209.82 | 211.02 | 211.02 207.92 | 209.82 | 101.14 | 101.72 | 101.72 | 101.14 100.23 |
| 1.5 | 0.5 | 103.78 | 103.98 | 103.98 103.46 | 103.78 | 100.39 | 100.58 | 100.58 | 100.39 100.08 |
| | 1.0 | 116.40 | 116.62 | 116.62 116.40 | 116.40 | 100.39 | 100.58 | 100.58 | 100.39 100.08 |
| | 1.5 | 137.43 | 137.70 | 137.70 137.00 | 137.43 | 100.39 | 100.58 | 100.58 | 100.39 100.08 |
| 2.0 | 0.5 | 101.53 | 101.59 | 101.59 101.42 | 101.53 | 100.13 | 100.19 | 100.19 | 100.03 100.03 |
| | 1.0 | 105.70 | 105.77 | 105.77 105.59 | 105.70 | 100.13 | 100.19 | 100.19 | 100.03 100.03 |
| | 1.5 | 112.65 | 112.72 | 112.72 112.54 | 112.65 | 100.13 | 100.19 | 100.19 | 100.03 100.03 |

# Empirical Study in Finite Correlation Coefficient in Two Phase Estimation

M. Khoshnevisan

Griffith University, Griffith Business School

Australia

F. Kaymarm

Massachusetts Institute of Technology

Department of Mechanical Engineering, USA

H. P. Singh, R Singh

Vikram University

Department of Mathematics and Statistics, India

F. Smarandache

University of New Mexico

Department of Mathematics, Gallup, USA.

**Abstract**

This paper proposes a class of estimators for population correlation coefficient when information about the population mean and population variance of one of the variables is not available but information about these parameters of another variable (auxiliary) is available, in two phase sampling and analyzes its properties. Optimum estimator in the class is identified with its variance formula. The estimators of the class involve unknown constants whose optimum values depend on unknown population parameters.Following (Singh, 1982) and (Srivastava and Jhajj, 1983), it has been shown that when these population parameters are replaced by their consistent estimates the resulting class of estimators has the same asymptotic variance as that of optimum estimator. An empirical study is carried out to demonstrate the performance of the constructed estimators.

**Keywords**: Correlation coefficient, Finite population, Auxiliary information, Variance.

**2000 MSC**: 92B28, 62P20

## 1. Introduction

Consider a finite population U= {1,2,..,i,..N}. Let y and x be the study and auxiliary variables taking values $y_i$ and $x_i$ respectively for the ith unit. The correlation coefficient between y and x is defined by

$$\rho_{yx} = S_{yx}/(S_y S_x) \qquad (1.1)$$

where

$$S_{yx} = (N-1)^{-1}\sum_{i=1}^{N}(y_i - \bar{Y})(x_i - \bar{X}), \ S_x^2 = (N-1)^{-1}\sum_{i=1}^{N}(x_i - \bar{X})^2, \ S_y^2 = (N-1)^{-1}\sum_{i=1}^{N}(y_i - \bar{Y})^2,$$

$$\bar{X} = N^{-1}\sum_{i=1}^{N}x_i, \ \bar{Y} = N^{-1}\sum_{i=1}^{N}y_i.$$

Based on a simple random sample of size $n$ drawn without replacement, $(x_i, y_i)$, i = 1,2,…,n; the usual estimator of $\rho_{yx}$ is the corresponding sample correlation coefficient :

$$r = s_{yx}/(s_x s_y) \qquad (1.2)$$

where $s_{yx} = (n-1)^{-1}\sum_{i=1}^{n}(y_i - \bar{y})(x_i - \bar{x}), \ s_x^2 = (n-1)^{-1}\sum_{i=1}^{n}(x_i - \bar{x})^2$

$$s_y^2 = (n-1)^{-1}\sum_{i=1}^{n}(y_i - \bar{y})^2, \ \bar{y} = n^{-1}\sum_{i=1}^{n}y_i, \ \bar{x} = n^{-1}\sum_{i=1}^{n}x_i.$$

The problem of estimating $\rho_{yx}$ has been earlier taken up by various authors including (Koop, 1970), (Gupta et. al., 1978, 1979), (Wakimoto, 1971), (Gupta and Singh, 1989), (Rana, 1989) and (Singh et. al., 1996) in different situations. (Srivastava and Jhajj, 1986) have further considered the problem of estimating $\rho_{yx}$ in the situations where the information on auxiliary variable *x* for all units in the population is available. In such situations, they have suggested a class of estimators for $\rho_{yx}$ which utilizes the known values of the population mean $\bar{X}$ and the population variance $S_x^2$ of the auxiliary variable *x*.

In this paper, using two – phase sampling mechanism, a class of estimators for $\rho_{yx}$ in the presence of the available knowledge ($\bar{Z}$ and $S_z^2$) on second auxiliary variable z is considered, when the population mean $\bar{X}$ and population variance $S_x^2$ of the main auxiliary variable *x* are not known.

## 2. The Suggested Class of Estimators

In many situations of practical importance, it may happen that no information is available on the population mean $\bar{X}$ and population variance $S_x^2$, we seek to estimate the population correlation coefficient $\rho_{yx}$ from a sample 's' obtained through a two-phase selection. Allowing simple random sampling without replacement scheme in each phase, the two- phase sampling scheme will be as follows:

(i) The first phase sample $s^*$ $(s^* \subset U)$ of fixed size $n_1$, is drawn to observe only *x* in order to furnish a good estimates of $\bar{X}$ and $S_x^2$.

(ii) Given $s^*$, the second- phase sample s $(s \subset s^*)$ of fixed size *n* is drawn to observe *y* only.

Let

$$\bar{x} = (1/n)\sum_{i \in s} x_i,\ \bar{y} = (1/n)\sum_{i \in s} y_i,\ \bar{x}^* = (1/n_1)\sum_{i \in s^*} x_i,\ s_x^2 = (n-1)^{-1}\sum_{i \in s}(x_i - \bar{x})^2,$$

$$s_x^{*2} = (n_1 - 1)^{-1}\sum_{i \in s^*}(x_i - \bar{x}^*)^2.$$

We write $u = \bar{x}/\bar{x}^*$, $v = s_x^2/s_x^{*2}$. Whatever be the sample chosen let $(u,v)$ assume values in a bounded closed convex subset, R, of the two-dimensional real space containing the point (1,1). Let *h* (*u, v*) be a function of *u* and *v* such that

$$h(1,1)=1 \qquad (2.1)$$

and such that it satisfies the following conditions:

1. The function $h(u,v)$ is continuous and bounded in R.
2. The first and second partial derivatives of $h(u,v)$ exist and are continuous and bounded in R.

Now one may consider the class of estimators of $\rho_{yx}$ defined by

$$\hat{\rho}_{hd} = r\, h(u,v) \qquad (2.2)$$

which is double sampling version of the class of estimators

$$\tilde{r}_t = r\, f(u^*, v^*)$$

Suggested by (Srivastava and Jhajj, 1986), where $u^* = \bar{x}/\bar{X}$, $v^* = s_x^2/S_x^2$ and $(\bar{X}, S_x^2)$ are known.

Sometimes even if the population mean $\bar{X}$ and population variance $S_x^2$ of $x$ are not known, information on a cheaply ascertainable variable $z$, closely related to $x$ but compared to $x$ remotely related to $y$, is available on all units of the population. This type of situation has been briefly discussed by, among others, (Chand, 1975), (Kiregyera, 1980, 1984).

Following (Chand, 1975) one may define a chain ratio-type estimator for $\rho_{yx}$ as

$$\hat{\rho}_{1d} = r\left(\frac{\bar{x}^*}{\bar{x}}\right)\left(\frac{\bar{Z}}{\bar{z}^*}\right)\left(\frac{s_x^{*2}}{s_x^2}\right)\left(\frac{S_z^2}{s_z^{*2}}\right) \qquad (2.3)$$

where the population mean $\bar{Z}$ and population variance $S_z^2$ of second auxiliary variable $z$ are known, and

$$\bar{z}^* = (1/n_1)\sum_{i\in s^*} z_i, \quad s_z^{*2} = (n_1-1)^{-1}\sum_{i\in s^*}(z_i - \bar{z}^*)^2$$

are the sample mean and sample variance of $z$ based on preliminary large sample $s^*$ of size $n_1\ (>n)$.

The estimator $\hat{\rho}_{1d}$ in (2.3) may be generalized as

$$\hat{\rho}_{2d} = r\left(\frac{\bar{x}}{\bar{x}^*}\right)^{\alpha_1}\left(\frac{s_x^2}{s_x^{*2}}\right)^{\alpha_2}\left(\frac{\bar{z}^*}{\bar{Z}}\right)^{\alpha_3}\left(\frac{s_z^{*2}}{S_z^2}\right)^{\alpha_4} \qquad (2.4)$$

where $\alpha_i$'s ($i=1,2,3,4$) are suitably chosen constants.

Many other generalization of $\hat{\rho}_{1d}$ is possible. We have, therefore, considered a more general class of $\rho_{yx}$ from which a number of estimators can be generated.

The proposed generalized estimators for population correlation coefficient $\rho_{yx}$ is defined by

$$\hat{\rho}_{td} = r\, t(u,v,w,a) \tag{2.5}$$

where $w = \bar{z}^*/\bar{Z}$, $a = s_z^{*2}/S_z^2$ and $t(u,v,w,a)$ is a function of $(u,v,w,a)$ such that

$$t(1,1,1,1) = 1 \tag{2.6}$$

Satisfying the following conditions:

(i) Whatever be the samples ($s^*$ and $s$) chosen, let $(u,v,w,a)$ assume values in a closed convex subset S, of the four dimensional real space containing the point $P=(1,1,1,1)$.

(ii) In S, the function $t(u,v,w,a)$ is continuous and bounded.

(iii) The first and second order partial derivatives of $t(u,v,w,a)$ exist and are continuous and bounded in S

To find the bias and variance of $\hat{\rho}_{td}$ we write

$$s_y^2 = S_y^2(1+e_1),\ \bar{x} = \bar{X}(1+e_1),\ \bar{x}^* = \bar{X}(1+e_1^*),\ s_x^2 = S_x^2(1+e_2)$$
$$s_x^{*2} = S_x^2(1+e_2^*),\ \bar{z}^* = \bar{Z}(1+e_3^*),\ s_z^{*2} = S_z^2(1+e_4^*),\ s_{yx} = S_{yx}(1+e_s)$$

such that $E(e_0) = E(e_1) = E(e_2) = E(e_5) = 0$ and $E(e_i^*) = 0\ \forall\ i = 1,2,3,4$,

and ignoring the finite population correction terms, we write to the first degree of approximation

$E(e_0^2) = (\delta_{400} - 1)/n$, $E(e_1^2) = C_x^2/n$, $E(e_1^{*2}) = C_x^2/n_1$, $E(e_2^2) = (\delta_{040} - 1)/n$,

$E(e_2^{*2}) = (\delta_{040} - 1)/n_1$, $E(e_3^{*2}) = C_z^2/n_1$, $E(e_4^{*2}) = (\delta_{004} - 1)/n_1$,

$E(e_5^2) = \{(\delta_{220}/\rho_{yx}^2) - 1\}/n$, $E(e_0 e_1) = \delta_{210} C_x/n$, $E(e_0 e_1^*) = \delta_{210} C_x/n_1$,

$E(e_0 e_2) = (\delta_{220} - 1)/n$, $E(e_0 e_2^*) = (\delta_{220} - 1)/n_1$, $E(e_0 e_3^*) = \delta_{201} C_z/n_1$,

$E(e_0 e_4^*) = (\delta_{202} - 1)/n_1$, $E(e_0 e_5) = \{(\delta_{310}/\rho_{yx}) - 1\}/n$,

$E(e_1 e_1^*) = C_x^2/n_1$, $E(e_1 e_2) = \delta_{030} C_x/n$, $E(e_1 e_2^*) = \delta_{030} C_x/n_1$,

$E(e_1 e_3^*) = \rho_{xz} C_x C_z/n_1$, $E(e_1 e_4^*) = \delta_{012} C_x/n_1$, $E(e_1 e_5) = (\delta_{120} C_x/\rho_{yx})/n$,

$E(e_1^* e_2) = \delta_{030} C_x/n_1$, $E(e_1^* e_2^*) = \delta_{030} C_x/n_1$, $E(e_1^* e_3^*) = \rho_{xz} C_x C_z/n_1$,

$E(e_1^* e_4^*) = \delta_{012} C_x/n_1$, $E(e_1^* e_5) = (\delta_{120} C_x/\rho_{yx})/n_1$,

$E(e_2 e_2^*) = (\delta_{040} - 1)/n_1$, $E(e_2 e_3^*) = \delta_{021} C_z/n_1$, $E(e_2 e_4^*) = (\delta_{022} - 1)/n_1$,

$E(e_2 e_5) = \{(\delta_{130}/\rho_{yx}) - 1\}/n$, $E(e_2^* e_3^*) = \delta_{021} C_z/n_1$,

$E(e_2^* e_4^*) = (\delta_{022} - 1)/n_1$, $E(e_2^* e_5) = \{(\delta_{130}/\rho_{yx}) - 1\}/n_1$,

$E(e_3^* e_4^*) = \delta_{003} C_z/n_1$, $E(e_3^* e_5) = (\delta_{111} C_z/\rho_{yx})/n_1$,

$E(e_4^* e_5) = \{(\delta_{112}/\rho_{yx}) - 1\}/n_1$.

where

$$\delta_{pqm} = \mu_{pqm}/(\mu_{200}^{p/2} \mu_{020}^{q/2} \mu_{002}^{m/2}), \quad \mu_{pqm} = (1/N)\sum_{i=1}^{N}(y_i - \overline{Y})^p (x_i - \overline{X})^q (z_i - \overline{Z})^m, \quad (p,q,m) \text{ being}$$

non-negative integers.

To find the expectation and variance of $\hat{\rho}_{td}$, we expand $t(u,v,w,a)$ about the point P = (1,1,1,1) in a second-order Taylor's series, express this value and the value of r in terms of e's. Expanding in powers of e's and retaining terms up to second power, we have

$$E(\hat{\rho}_{td}) = \rho_{yx} + o(n^{-1}) \tag{2.7}$$

which shows that the bias of $\hat{\rho}_{td}$ is of the order $n^{-1}$ and so up to order $n^{-1}$, mean square error and the variance of $\hat{\rho}_{td}$ are same.

Expanding $(\hat{\rho}_{td} - \rho_{yx})^2$, retaining terms up to second power in e's, taking expectation and using the above expected values, we obtain the variance of $\hat{\rho}_{td}$ to the first degree of approximation, as

$$Var(\hat{\rho}_{td}) = Var(r) + (\rho_{yx}^2/n)[C_x^2 t_1^2(P) + (\delta_{040}-1)t_2^2(P) - At_1(P) - Bt_2(P) + 2\delta_{030}C_x t_1(P)t_2(P)]$$
$$- (\rho_{yx}^2/n_1)[C_x^2 t_1^2(P) + (\delta_{040}-1)t_2^2(P) - C_z^2 t_3^2(P) - (\delta_{004}-1)t_4^2(P) - At_1(P) -$$
$$Bt_2(P) + Dt_3(P) + Ft_4(P) + 2\delta_{030}C_x t_1(P)t_2(P) - 2\delta_{003}C_z t_3(P)t_4(P)]$$

(2.8)

where $t_1(P)$, $t_2(P)$, $t_3(P)$ and $t_4(P)$ respectively denote the first partial derivatives of $t(u,v,w,a)$ white respect to $u,v,w$ and $a$ respectively at the point $P = (1,1,1,1)$,

$$Var(r) = (\rho_{yx}^2/n)[(\delta_{220}/\rho_{yx}^2) + (1/4)(\delta_{040} + \delta_{400} + 2\delta_{220}) - \{(\delta_{130} + \delta_{310})/\rho_{yx}\}] \quad (2.9)$$

$$A = \{\delta_{210} + \delta_{030} - 2(\delta_{120}/\rho_{yx})\}C_x, B = \{\delta_{220} + \delta_{040} - 2(\delta_{130}/\rho_{yx})\},$$
$$D = \{\delta_{201} + \delta_{021} - 2(\delta_{111}/\rho_{yx})\}C_z, F = \{\delta_{202} + \delta_{022} - 2(\delta_{112}/\rho_{yx})\}$$

Any parametric function $t(u,v,w,a)$ satisfying (2.6) and the conditions (1) and (2) can generate an estimator of the class (2.5).

The variance of $\hat{\rho}_{td}$ at (2.6) is minimized for

$$\left. \begin{array}{l} t_1(P) = \dfrac{[A(\delta_{040}-1) - B\delta_{030}C_x]}{2C_x^2(\delta_{040} - \delta_{030}^2 - 1)} = \alpha \text{(say)}, \\[2mm] t_2(P) = \dfrac{(BC_x^2 - A\delta_{030}C_x)}{2C_x^2(\delta_{040} - \delta_{030}^2 - 1)} = \beta \text{(say)}, \\[2mm] t_3(P) = \dfrac{[D(\delta_{004}-1) - F\delta_{003}C_z]}{2C_z^2(\delta_{004} - \delta_{030}^2 - 1)} = \gamma \text{(say)}, \\[2mm] t_4(P) = \dfrac{(C_z^2 F - D\delta_{003}C_z)}{2C_z^2(\delta_{004} - \delta_{003}^2 - 1)} = \delta \text{(say)}, \end{array} \right\} \quad (2.10)$$

Thus the resulting (minimum) variance of $\hat{\rho}_{td}$ is given by

$$\min. Var(\hat{\rho}_{td}) = Var(r) - \left(\frac{1}{n} - \frac{1}{n_1}\right)\rho_{yx}^2\left[\frac{A^2}{4C_x^2} + \frac{\{(A/C_x)\delta_{030} - B\}^2}{4(\delta_{040} - \delta_{030}^2 - 1)}\right]$$
$$- (\rho_{yx}^2/n_1)\left[\frac{D^2}{4C_z^2} + \frac{\{(D/C_z)\delta_{003} - F\}^2}{4(\delta_{004} - \delta_{003}^2 - 1)}\right]$$

(2.11)

It is observed from (2.11) that if optimum values of the parameters given by (2.10) are used, the variance of the estimator $\hat{\rho}_{td}$ is always less than that of r as the last two terms on the right hand sides of (2.11) are non-negative.

Two simple functions $t(u,v,w,a)$ satisfying the required conditions are

$$t(u,v,w,a) = 1 + \alpha_1(u-1) + \alpha_2(v-1) + \alpha_3(w-1) + \alpha_4(a-1)$$

$$t(u,v,w,a) = u^{\alpha_1} v^{\alpha_2} w^{\alpha_3} a^{\alpha_4}$$

and for both these functions $t_1(P) = \alpha_1$, $t_2(P) = \alpha_2$, $t_3(P) = \alpha_3$ and $t_4(P) = \alpha_4$. Thus one should use optimum values of $\alpha_1, \alpha_2, \alpha_3$ and $\alpha_4$ in $\hat{\rho}_{td}$ to get the minimum variance. It is to be noted that the estimated $\hat{\rho}_{td}$ attained the minimum variance only when the optimum values of the constants $\alpha_i$ (i=1,2,3,4), which are functions of unknown population parameters, are known. To use such estimators in practice, one has to use some guessed values of population parameters obtained either through past experience or through a pilot sample survey. It may be further noted that even if the values of the constants used in the estimator are not exactly equal to their optimum values as given by (2.8) but are close enough, the resulting estimator will be better than the conventional estimator, as has been illustrated by (Das and Tripathi, 1978, Sec.3).

If no information on second auxiliary variable z is used, then the estimator $\hat{\rho}_{td}$ reduces to $\hat{\rho}_{hd}$ defined in (2.2). Taking $z \equiv 1$ in (2.8), we get the variance of $\hat{\rho}_{hd}$ to the first degree of approximation, as

$$Var(\hat{\rho}_{hd}) = Var(r) + \left(\frac{1}{n} - \frac{1}{n_1}\right)\rho_{yx}^2 \left[C_x^2 h_1^2(1,1) + (\delta_{040}-1)h_2^2(1,1) - Ah_1(1,1) - Bh_2(1,1) + 2\delta_{030}C_x h_1(1,1)h_2(1,1)\right]$$

(2.12)

which is minimized for

$$h_1(1,1) = \frac{[A(\delta_{040}-1) - B\delta_{030}C_x]}{2C_x^2(\delta_{040} - \delta_{030}^2 - 1)}, \quad h_2(1,1) = \frac{(BC_x^2 - A\delta_{030}C_x)}{2C_x^2(\delta_{040} - \delta_{030}^2 - 1)} \quad (2.13)$$

Thus the minimum variance of $\hat{\rho}_{hd}$ is given by

$$\min.\text{Var}(\hat{\rho}_{hd}) = \text{Var}(r) - \left(\frac{1}{n} - \frac{1}{n_1}\right) \rho_{yx}^2 \left[\frac{A^2}{4C_x^2} + \frac{\{(A/C_x)\delta_{030} - B\}^2}{4(\delta_{040} - \delta_{030}^2 - 1)}\right] \quad (2.14)$$

It follows from (2.11) and (2.14) that

$$\min.\text{Var}(\hat{\rho}_{td}) - \min.\text{Var}(\hat{\rho}_{hd}) = \left(\rho_{yx}^2/n_1\right) \left[\frac{D^2}{4C_z^2} + \frac{\{(D/C_z)\delta_{003} - F\}^2}{4(\delta_{004} - \delta_{003}^2 - 1)}\right] \quad (2.15)$$

which is always positive. Thus the proposed estimator $\hat{\rho}_{td}$ is always better than $\hat{\rho}_{hd}$.

## 3. A Wider Class of Estimators

In this section we consider a class of estimators of $\rho_{yx}$ wider than (2.5) given by

$$\hat{\rho}_{gd} = g(r,u,v,w,a) \quad (3.1)$$

where $g(r,u,v,w,a)$ is a function of $r,u,v,w,a$ and such that

$g(\rho,1,1,1,1) = \hat{\rho}_{td}$ and $\left[\frac{\partial g(\cdot)}{\partial r}\right]_{(\rho,1,1,1)} = 1$

Proceeding as in section 2, it can easily be shown, to the first order of approximation, that the minimum variance of $\hat{\rho}_{gd}$ is same as that of $\hat{\rho}_{td}$ given in (2.11).

It is to be noted that the difference-type estimator

$r_d = r + \alpha_1 (u-1) + \alpha_2 (v-1) + \alpha_3 (w-1) + \alpha_4 (a-1)$, is a particular case of $\hat{\rho}_{gd}$, but it is not the member of $\hat{\rho}_{td}$ in (2.5).

## 4. Optimum Values and Their Estimates

The optimum values $t_1(P) = \alpha$, $t_2(P) = \beta$, $t_3(P) = \gamma$ and $t_4(P) = \delta$ given at (2.10) involves unknown population parameters. When these optimum values are substituted in (2.5), it no longer remains an estimator since it involves unknown ($\alpha, \beta, \gamma, \delta$), which are functions of unknown population parameters, say, $\delta_{pqm}$ (p, q, m= 0,1,2,3,4), $C_x$, $C_z$ and $\rho_{yx}$ itself. Hence it is advisable to replace them by their consistent estimates from sample values. Let ($\hat{\alpha}, \hat{\beta}, \hat{\gamma}, \hat{\delta}$) be consistent estimators of $t_1(P), t_2(P)$, $t_3(P)$ and $t_4(P)$ respectively, where

$$\hat{t}_1(P) = \hat{\alpha} = \frac{[\hat{A}(\hat{\delta}_{040} - 1) - \hat{B}\hat{\delta}_{030}\hat{C}_x]}{2\hat{C}_x^2(\hat{\delta}_{040} - \hat{\delta}_{030}^2 - 1)}, \qquad \hat{t}_2(P) = \hat{\beta} = \frac{[\hat{B}\hat{C}_x^2 - \hat{A}\hat{\delta}_{030}\hat{C}_x]}{2\hat{C}_x^2(\hat{\delta}_{040} - \hat{\delta}_{030}^2 - 1)},$$

$$\hat{t}_3(P) = \hat{\gamma} = \frac{[\hat{D}(\hat{\delta}_{004} - 1) - \hat{F}\hat{\delta}_{003}\hat{C}_z]}{2\hat{C}_z^2(\hat{\delta}_{004} - \hat{\delta}_{003}^2 - 1)}, \qquad \hat{t}_4(P) = \hat{\delta} = \frac{[\hat{C}_z^2\hat{F} - \hat{D}\hat{\delta}_{003}\hat{C}_z]}{2\hat{C}_z^2(\hat{\delta}_{004} - \hat{\delta}_{003}^2 - 1)},$$

(4.1)

with

$$\hat{A} = [\hat{\delta}_{210} + \hat{\delta}_{030} - 2(\hat{\delta}_{120}/r)]\hat{C}_x, \qquad \hat{B} = [\hat{\delta}_{220} + \hat{\delta}_{040} - 2(\hat{\delta}_{130}/r)],$$

$$\hat{D} = [\hat{\delta}_{201} + \hat{\delta}_{021} - 2(\hat{\delta}_{111}/r)]\hat{C}_z, \qquad \hat{F} = [\hat{\delta}_{202} + \hat{\delta}_{022} - 2(\hat{\delta}_{112}/r)],$$

$$\hat{C}_x = s_x/\bar{x}, \ \hat{C}_z = s_z/\bar{z}, \ \hat{\delta}_{pqm} = \hat{\mu}_{pqm}/\left(\hat{\mu}_{200}^{p/2} \hat{\mu}_{020}^{q/2} \hat{\mu}_{002}^{m/2}\right)$$

$$\hat{\mu}_{pqm} = (1/n)\sum_{i=1}^{n}(y_i - \bar{y})^p (x_i - \bar{x})^q (z_i - \bar{z})^m$$

$$\bar{z} = (1/n)\sum_{i=1}^{n} z_i, \ s_x^2 = (n-1)^{-1}\sum_{i=1}^{n}(x_i - \bar{x})^2, \ \bar{x} = (1/n)\sum_{i=1}^{n} x_i,$$

$$r = s_{yx}/(s_y s_x), \ s_y^2 = (n-1)^{-1}\sum_{i=1}^{n}(y_i - \bar{y})^2, \ s_z^2 = (n-1)^{-1}\sum_{i=1}^{n}(x_i - \bar{z})^2.$$

We then replace $(\alpha, \beta, \gamma, \delta)$ by $(\hat{\alpha}, \hat{\beta}, \hat{\gamma}, \hat{\delta})$ in the optimum $\hat{\rho}_{td}$ resulting in the estimator $\hat{\rho}_{td}^*$ say, which is defined by

$$\hat{\rho}_{td}^* = rt^*(u, v, w, a, \hat{\alpha}, \hat{\beta}, \hat{\gamma}, \hat{\delta}), \qquad (4.2)$$

*where the function* t*(U), U= $(u, v, w, a, \hat{\alpha}, \hat{\beta}, \hat{\gamma}, \hat{\delta})$ *is derived from the the function* t(u,v,w,a) *given at (2.5) by replacing the unknown constants involved in it by the consistent estimates of optimum values. The condition (2.6) will then imply that*

$$t^*(P^*) = 1 \qquad (4.3)$$

where $P^* = (1,1,1,1, \alpha, \beta, \gamma, \delta)$

We further assume that

$$t_1(P^*) = \frac{\partial t^*(U)}{\partial u}\bigg]_{U=P^*} = \alpha, \qquad\qquad t_2(P^*) = \frac{\partial t^*(U)}{\partial v}\bigg]_{U=P^*} = \beta$$

$$t_3(P^*) = \frac{\partial t^*(U)}{\partial w}\bigg]_{U=P^*} = \gamma, \qquad\qquad t_4(P^*) = \frac{\partial t^*(U)}{\partial a}\bigg]_{U=P^*} = \delta \qquad (4.4)$$

$$t_5(P^*) = \frac{\partial t^*(U)}{\partial \hat{\alpha}}\bigg]_{U=P^*} = o \qquad\qquad t_6(P^*) = \frac{\partial t^*(U)}{\partial \hat{\beta}}\bigg]_{U=P^*} = o$$

$$t_7(P^*) = \frac{\partial t^*(U)}{\partial \hat{\gamma}}\bigg]_{U=P^*} = o \qquad\qquad t_8(P^*) = \frac{\partial t^*(U)}{\partial \hat{\delta}}\bigg]_{U=P^*} = o$$

Expanding $t^*(U)$ about $P^* = (1,1,1,1, \alpha, \beta, \gamma, \delta)$, in Taylor's series, we have

$$\hat{\rho}_{td}^* = r[t^*(P^*) + (u-1)t_1^*(P^*) + (v-1)t_2^*(P^*) + (w-1)t_3^*(P^*) + (a-1)t_4^*(P^*) + (\hat{\alpha}-\alpha)t_5^*(P^*)$$
$$+ (\hat{\beta}-\beta)t_6^*(P^*) + (\hat{\gamma}-\gamma)t_7^*(P^*) + (\hat{\delta}-\delta)t_8^*(P^*) + \text{second order terms}]$$

(4.5)

Using (4.4) in (4.5) we have

$$\hat{\rho}_{td}^* = r[1 + (u-1)\alpha + (v-1)\beta + (w-1)\gamma + (a-1)\delta + \text{second order terms}] \qquad (4.6)$$

Expressing (4.6) in term of e's squaring and retaining terms of e's up to second degree, we have

$$(\hat{\rho}_{td}^* - \rho_{yx})^2 = \rho_{yx}^2 [\frac{1}{2}(2e_5 - e_0 - e_2) + \alpha(e_1 - e_1^*) + \beta(e_2 - e_2^*) + \gamma e_3^* + \delta e_4^*]^2 \qquad (4.7)$$

Taking expectation of both sides in (4.7), we get the variance of $\hat{\rho}_{td}^*$ to the first degree of approximation, as

$$Var(\hat{\rho}_{td}^*) = Var(r) - (\frac{1}{n} - \frac{1}{n_1})\rho_{yx}^2 \left[ \frac{A^2}{4C_x^2} + \frac{\{(A/C_x)\delta_{030} - B\}^2}{4(\delta_{040} - \delta_{030}^2 - 1)} \right]$$
$$+ (\rho_{yx}^2 / n_1)\left[ \frac{D^2}{4C_z^2} + \frac{\{(D/C_z)\delta_{003} - F\}^2}{4(\delta_{004} - \delta_{003}^2 - 1)} \right] \quad (4.8)$$

which is same as (2.11), we thus have established the following result.

**Result 4.1:** If optimum values of constants in (2.10) are replaced by their consistent estimators and conditions (4.3) and (4.4) hold good, the resulting estimator $\hat{\rho}_{td}^*$ has the same variance to the first degree of approximation, as that of optimum $\hat{\rho}_{td}$.

**Remark 4.1:** It may be easily examined that some special cases:

(i) $\hat{\rho}_{td1}^* = r u^{\hat{\alpha}} v^{\hat{\beta}} w^{\hat{\gamma}} a^{\hat{\delta}}$, (ii) $\hat{\rho}_{td2}^* = r \frac{\{1 + \hat{\alpha}(u-1) + \hat{\gamma}(w-1)\}}{\{1 - \hat{\beta}(v-1) - \hat{\delta}(a-1)\}}$

(iii) $\hat{\rho}_{td3}^* = r[1 + \hat{\alpha}(u-1) + \hat{\beta}(u-1) + \hat{\gamma}(w-1) + \hat{\delta}(a-1)]$

(iv) $\hat{\rho}_{td4}^* = r[1 - \hat{\alpha}(u-1) - \hat{\beta}(u-1) - \hat{\gamma}(w-1) - \hat{\delta}(a-1)]^{-1}$

of $\hat{\rho}_{td}^*$ satisfy the conditions (4.3) and (4.4) and attain the variance (4.8).

**Remark 4.2:** The efficiencies of the estimators discussed in this paper can be compared for fixed cost, following the procedure given in (Sukhatme et. al., 1984).

**5. Empirical Study**

To illustrate the performance of various estimators of population correlation coefficient, we consider the data given in (Murthy, 1967, p. 226]. The variates are:

y=output, x=Number of Workers, z =Fixed Capital

N=80, n=10, $n_1$=25,

$\bar{X} = 283.875$, $\bar{Y} = 5182.638$, $\bar{Z} = 1126$, $C_x = 0.9430$, $C_y = 0.3520$, $C_z = 0.7460$,

$\delta_{003} = 1.030$, $\delta_{004} = 2.8664$, $\delta_{021} = 1.1859$, $\delta_{022} = 3.1522$, $\delta_{030} = 1.295$, $\delta_{040} = 3.65$,

$\delta_{102} = 0.7491$, $\delta_{120} = 0.9145$, $\delta_{111} = 0.8234$, $\delta_{130} = 2.8525$,

$\delta_{112} = 2.5454$, $\delta_{210} = 0.5475$, $\delta_{220} = 2.3377$, $\delta_{201} = 0.4546$, $\delta_{202} = 2.2208$, $\delta_{300} = 0.1301$,

$\delta_{400} = 2.2667$, $\rho_{yx} = 0.9136$, $\rho_{xz} = 0.9859$, $\rho_{yz} = 0.9413$.

The percent relative efficiencies (PREs) of $\hat{\rho}_{1d}, \hat{\rho}_{hd}, \hat{\rho}_{td}$ with respect to conventional estimator r have been computed and compiled in Table 5.1.

**Table 5.1: The PRE's of different estimators of $\rho_{yx}$**

| Estimator | r | $\hat{\rho}_{hd}$ | $\hat{\rho}_{td}$ (or $\hat{\rho}_{td}^*$) |
|---|---|---|---|
| PRE(.,r) | 100 | 129.147 | 305.441 |

Table 5.1 clearly shows that the proposed estimator $\hat{\rho}_{td}$ (or $\hat{\rho}_{td}^*$) is more efficient than $r$ and $\hat{\rho}_{hd}$.

**MASS – *Modified Assignment* Algorithm in Facilities Layout Planning**

Dr. Sukanto Bhattacharya
Department of Business Administration
Alaska Pacific University, AK 99508, USA

Dr. Florentin Smarandache
University of New Mexico
200 College Road, Gallup, USA

Dr. M. Khoshnevisan
School of Accounting & Finance
Griffith University, Australia

**Abstract**

In this paper we have proposed a semi-heuristic optimization algorithm for designing optimal plant layouts in process-focused manufacturing/service facilities. Our proposed algorithm marries the well-known CRAFT (Computerized Relative Allocation of Facilities Technique) with the Hungarian assignment algorithm. Being a semi-heuristic search, our algorithm is likely to be more efficient in terms of computer CPU engagement time as it tends to converge on the global optimum faster than the traditional CRAFT algorithm - a pure heuristic. We also present a numerical illustration of our algorithm.

**Key Words**: Facilities layout planning, load matrix, CRAFT, Hungarian assignment algorithm

**Introduction**

The fundamental integration phase in the design of productive systems is the layout of production facilities. A working definition of layout may be given as the arrangement of machinery and flow of materials from one facility to another, which minimizes material-handling costs while considering any physical restrictions on such arrangement. Usually this layout design is either on considerations of machine-time cost and product availability; thereby making the production system *product-focused*; or on considerations of quality and flexibility; thereby making the production system *process-focused*. It is natural that while *product-focused* systems are better off with a 'line layout' dictated by available technologies and prevailing job designs, *process-focused* systems, which are more concerned with job organization, opt for a 'functional layout'. Of course, in reality the actual facility layout often lies somewhere in between a pure line layout and a pure functional layout format; governed by the specific demands of a particular production plant. Since our present paper concerns only functional layout design for process-focused systems, this is the only layout design we will discuss here.

The main goal to keep in mind is to minimize material handling costs - therefore the departments that incur the most interdepartmental movement should be located closest to one another. The main type of design layouts is *Block diagramming*, which refers to the movement of materials in existing or proposed facility. This information is usually provided with a *from/to* chart or *load summary* chart, which gives the average number of units loads moved between departments. A load-unit can be a single unit, a pallet of material, a bin of material, or a crate of material. The next step is to design the layout by calculating the composite movements between departments and rank them from most movement to least movement. Composite movement refers to the back-and-forth movement between each pair of departments. Finally, trial layouts are placed on a grid that graphically represents the relative distances between departments. This grid then becomes the objective of optimization when determining the optimal plant layout.

We give a visual representation of the basic operational considerations in a *process-focused* system schematically as follows:

**Figure 1**

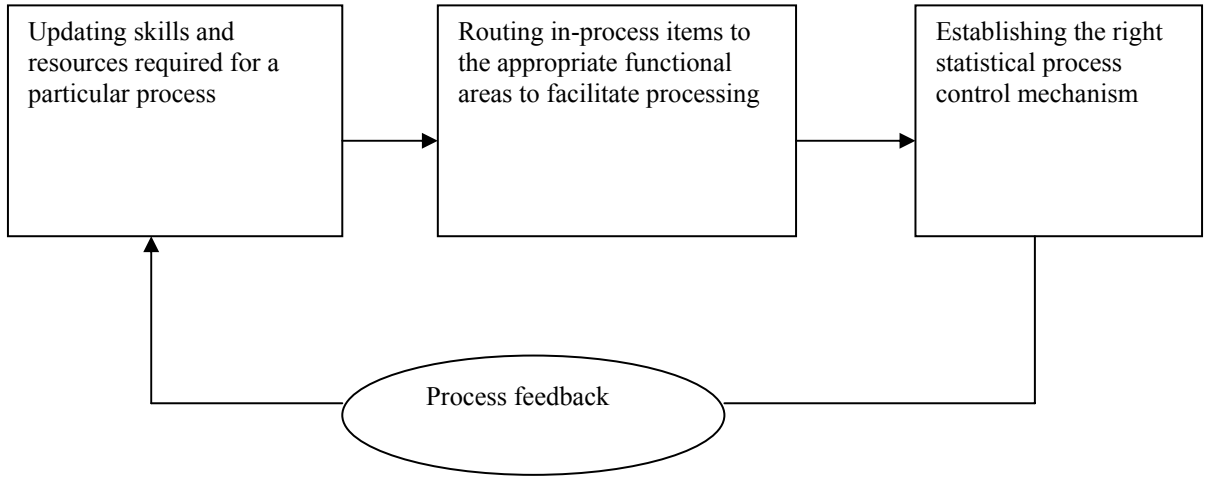

In designing the optimal functional layout, the fundamental question to be addressed is that of 'relative location of facilities'. The locations will depend on the need for one pair of facilities to be adjacent (or physically close) to each other relative to the need for all other pairs of facilities to be similarly adjacent (or physically close) to each other. Locations must be allocated based on the relative gains and losses for the alternatives and seek to minimize some indicative measure of the cost of having non-adjacent locations of facilities. Constraints of space prevents us from going into the details of the several criteria used to determine the gains or losses from the relative location of facilities and the available sequence analysis techniques for addressing the question; for which we refer the interested reader to any standard handbook of production/operations management.

**Computerized Relative Allocation of Facilities Technique (CRAFT)**

CRAFT (Buffa, Armour and Vollman, 1964) is a computerized heuristic algorithm that takes in *load matrix* of interdepartmental flow and transaction costs with a representation of a block layout as the inputs. The block layout could either be an existing layout or; for a new facility, any arbitrary initial layout. The algorithm then computes the departmental locations and returns an estimate of the total interaction costs for the initial layout. The governing algorithm is designed to compute the impact on a cost measure for two-way or three-way swapping in the location of the facilities. For each swap, the various interaction costs are computed afresh and the load matrix and the change in cost (increase or decrease) is noted and stored in the RAM. The algorithm proceeds this way through all possible combinations of swaps accommodated by the software. The basic procedure is repeated a number of times resulting in a more efficient block layout every time till such time when no further cost reduction is possible. The final block layout is then printed out to serve as the basis for a detailed layout template of the facilities at a later stage. Since its formulation, more powerful versions of CRAFT have been developed but these too follow the same, basic heuristic routine and therefore tend to be highly CPU-intensive (Khalil, 1973; Hicks and Cowan, 1976).

The basic computational disadvantage of a CRAFT-type technique is that one always has got to start with an arbitrary initial solution (Carrie, 1980). This means that there is no mathematical certainty of attaining the desired optimal solution after a given number of iterations. If the starting solution is quite close to the optimal solution by chance, then the final solution is attained only after a few iterations. However, as there is no guarantee that the starting solution will be close to the global optimum, the expected number of iterations required to arrive at the final solution tend to be quite large thereby straining computing resources (Driscoll and Sangi, 1988).

In our present paper we propose and illustrate the Modified Assignment (MASS) algorithm as an extension to the traditional CRAFT, to enable faster convergence to the optimal solution. This we propose to do by marrying CRAFT technique with the Hungarian assignment algorithm. As our proposed algorithm is semi-heuristic, it is likely to be less CPU-intensive than any traditional, purely heuristic CRAFT-type algorithm.

**The Hungarian assignment algorithm**

A general assignment problem may be framed as a special case of the *balanced transportation problem* with availability and demand constraints summing up to unity. Mathematically, it has the following *general linear programming* form:

$$\text{Minimize } \Sigma\Sigma\ C_{ij}X_{ij}$$
$$\text{Subject to } \Sigma X_{ij} = 1, \text{ for each } i, j = 1, 2 \ldots n.$$

In words, the problem may be stated as assigning each of n individuals to n jobs so that exactly one individual is assigned to each job in such a way as to minimize the total cost.

To ensure satisfaction of the basic requirements of the assignment problem, the basic feasible solutions of the corresponding balanced transportation problem must be integer valued. However, any such basic feasible solution will contain (2n – 1) variables out of which (n – 1) variables will be zero thereby introducing a high level of degeneracy in the solution making the usual solution technique of a transportation problem very inefficient.

This has resulted in mathematicians devising an alternative, more efficient algorithm for solving this class of problems, which has come to be commonly known as the *Hungarian assignment algorithm*. Basically, this algorithm draws from a simple theorem in linear algebra which says that if a constant number is added to any row and/or column of the cost matrix of an assignment-type problem, then the resulting assignment-type problem has exactly the same set of optimal solutions as the original problem and vice versa.

**Proof:**

Let $A_i$ and $B_j$ (i, j = 1, 2 … n) be added to the ith row and/or jth column respectively of the cost matrix. Then the revised cost elements are $C_{ij}^* = C_{ij} + A_i + B_j$. The revised cost of assignment is $\Sigma\Sigma C_{ij}^* X_{ij} = \Sigma\Sigma\ (C_{ij} + A_i + B_j)\ X_{ij} = \Sigma\Sigma C_{ij}X_{ij} + \Sigma A_i\ \Sigma X_{ij} + \Sigma B_j \Sigma X_{ij}$. But by the imposed assignment constraint $\Sigma X_{ij} = 1$ **(for i, j = 1, 2 … n)**, we have the revised cost as $\Sigma\Sigma C_{ij}X_{ij} + \Sigma A_i + \Sigma B_j$ i.e. the cost differs from the original by a constant. As the revised costs differ from the originals by a constant, which is independent of the decision variables, an optimal solution to one is also optimal solution to the other and vice versa.

This theorem can be used in two different ways to solve the assignment problem. First, if in an assignment problem, some cost elements are negative, the problem may be converted into an equivalent assignment problem by adding a positive constant to each of the entries in the cost matrix so that they all become non-negative. Next, the important thing to look for is a feasible solution that has zero assignment cost after adding suitable constants to the rows and columns. Since it has been assumed that all entries are now non-negative, this assignment must be the globally optimal one (Mustafi, 1996).

Given a zero assignment, a straight line is drawn through it (a horizontal line in case of a row and a vertical line in case of a column), which prevents any other assignment in that particular row/column. The governing algorithm then seeks to find the minimum number of such straight lines, which would cover all the zero entries to avoid any redundancy. Let us say that *k* such lines are required to cover all the zeroes. Then the *necessary condition* for optimality is that number of zeroes assigned is equal to *k* and the *sufficient condition* for optimality is that *k* is equal to *n* for an *n x n* cost matrix.

**The MASS (Modified Assignment) algorithm**

The basic idea of our proposed algorithm is to develop a systematic scheme to arrive at the initial input block layout to be fed into the CRAFT program so that the program does not have to start off from any initial (and possibly inefficient) solution. Thus, by subjecting the problem of finding an initial block layout to a mathematical scheme, we in effect reduce the purely heuristic algorithm of CRAFT to a semi-heuristic one. Our proposed MASS algorithm follows the following *sequential* steps:

**Step 1:** We formulate the load matrix such that each entry $l_{ij}$ represents the load carried from facility i to facility j

**Step 2:** We insert $l_{ij} = M$, where M is a large positive number, into all the vacant cells of the load matrix signifying that no inter-facility load transportation is required or possible between the i$^{th}$ and j$^{th}$ vacant cells

**Step 3:** We solve the problem on the lines of a standard assignment problem using the Hungarian assignment algorithm treating the load matrix as the cost matrix

**Step 4:** We draft the initial block layout trying to keep the inter-facility distance $d_{ij}^*$ between the i$^{th}$ and j$^{th}$ *assigned* facilities to the minimum possible magnitude, subject to the available floor area and architectural design of the shop floor

**Step 5:** We proceed using the CRAFT program to arrive at the optimal layout by iteratively improving upon the starting solution provided by the Hungarian assignment algorithm till the overall load function **L = ΣΣ $l_{ij}d_{ij}^*$** subject to any particular bounds imposed on the problem

The Hungarian assignment algorithm will ensure that the initial block layout is at least very close to the global optimum if not globally optimal itself. Therefore the subsequent CRAFT procedure will converge on the global optimum much faster starting from this near-optimal initial input block layout and will be much less CPU-intensive that any traditional CRAFT-type algorithm. Thus MASS is not a stand-alone optimization tool but rather a rider on the traditional CRAFT that tries to ensure faster convergence to the optimal block layout for process-focused systems, by making the search semi-heuristic.

We provide a numerical illustration of the MASS algorithm in the Appendix by designing the optimal block layout of a small, single-storied, process-focused manufacturing plant with six different facilities and a rectangular shop floor design. The model can however be extended to cover bigger plants with more number of facilities. Also the MASS approach we have advocated here can even be extended to deal with the multi-floor version of CRAFT (Johnson, 1982) by constructing a separate assignment table for each floor subject to any predecessor-successor relationship among the facilities.

---

**Appendix: Numerical illustration of MASS**

We consider a small, single-storied process-focused manufacturing plant with a rectangular shop floor plan having six different facilities. We mark these facilities as $F_I$, $F_{II}$, $F_{III}$, $F_{IV}$, $F_V$ and $F_{VI}$. The architectural design requires that there be an aisle of at least 2 meters width between two adjacent facilities and the total floor area of the plant is 64 meters x 22 meters. Based on the different types of jobs processed, the loads to be transported between the different facilities are supplied in the following load matrix:

**Table 1**

|  | $F_I$ | $F_{II}$ | $F_{III}$ | $F_{IV}$ | $F_V$ | $F_{VI}$ |
|---|---|---|---|---|---|---|
| $F_I$ | – | 20 | – | – | – | 25 |
| $F_{II}$ | 10 | – | 15 | – | – | – |
| $F_{III}$ | – | – | – | 30 | – | – |
| $F_{IV}$ | – | – | 50 | – | – | 40 |
| $F_V$ | – | – | – | – | – | 10 |
| $F_{VI}$ | – | – | – | – | 15 | – |

We put in a very large positive value M in each of the vacant cells of the load matrix to signify that no inter-facility transfer of load is required or is permissible for these cells:

**Table 2**

|          | $F_I$ | $F_{II}$ | $F_{III}$ | $F_{IV}$ | $F_V$ | $F_{VI}$ |
|----------|-------|----------|-----------|----------|-------|----------|
| $F_I$    | M     | 20       | M         | M        | M     | 25       |
| $F_{II}$ | 10    | M        | 15        | M        | M     | M        |
| $F_{III}$| M     | M        | M         | 30       | M     | M        |
| $F_{IV}$ | M     | M        | 50        | M        | M     | 40       |
| $F_V$    | M     | M        | M         | M        | M     | 10       |
| $F_{VI}$ | M     | M        | M         | M        | 15    | M        |

Next we apply the standard Hungarian assignment algorithm to obtain the initial solution:

Assignment table after first iteration:

**Table 3**

|          | $F_I$  | $F_{II}$ | $F_{III}$ | $F_{IV}$ | $F_V$  | $F_{VI}$ |
|----------|--------|----------|-----------|----------|--------|----------|
| $F_I$    | *M-20* | *0*      | M-25      | *M-20*   | *M-20* | *5*      |
| $F_{II}$ | *0*    | *M-10*   | *0*       | *M-10*   | *M-10* | *M-10*   |
| $F_{III}$| M-30   | M-30     | M-35      | *0*      | M-30   | M-30     |
| $F_{IV}$ | M-40   | M-40     | 5         | *M-40*   | M-40   | *0*      |
| $F_V$    | M-10   | M-10     | M-15      | *M-10*   | M-10   | *0*      |
| $F_{VI}$ | M-15   | M-15     | M-20      | *M-15*   | *0*    | *M-15*   |

There are two rows and three columns that are covered i.e. $k = 5$. But as this is a 6x6 load matrix, the above solution is sub-optimal. So we make a second iteration:

**Table 4**

|      | $F_I$ | $F_{II}$ | $F_{III}$ | $F_{IV}$ | $F_V$ | $F_{VI}$ |
|------|-------|----------|-----------|----------|-------|----------|
| $F_I$ | *M-20* | *0* | M-25 | *M-15* | *M-15* | *10* |
| $F_{II}$ | *0* | M-10 | *0* | *M-5* | M-5 | *M-5* |
| $F_{III}$ | *M-35* | M-35 | *M-40* | *0* | M-30 | *M-30* |
| $F_{IV}$ | *M-45* | M-45 | *0* | *M-40* | M-40 | *0* |
| $F_V$ | *M-15* | M-15 | *M-20* | *M-10* | M-10 | *0* |
| $F_{VI}$ | *M-20* | *M-20* | *M-25* | *M-15* | *0* | *M-15* |

Now columns $F_I$, $F_{III}$, $F_{IV}$, $F_{VI}$ and rows $F_I$ and $F_{VI}$ are covered i.e. k = 6. As this is a 6x6 load matrix the above solution is optimal. The optimal assignment table (subject to the 2 meters of aisle between adjacent facilities) is shown below:

**Table 5**

|      | $F_I$ | $F_{II}$ | $F_{III}$ | $F_{IV}$ | $F_V$ | $F_{VI}$ |
|------|-------|----------|-----------|----------|-------|----------|
| $F_I$ | – | * | – | – | – | – |
| $F_{II}$ | * | – | – | – | – | – |
| $F_{III}$ | – | – | – | * | – | – |
| $F_{IV}$ | – | – | * | – | – | – |
| $F_V$ | – | – | – | – | – | * |
| $F_{VI}$ | – | – | – | – | * | – |

Initial layout of facilities as dictated by the Hungarian assignment algorithm:

**Figure 2**

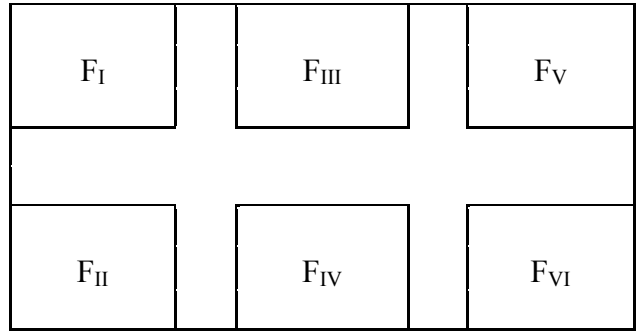

The above layout conforms to the rectangular floor plan of the plant and also places the assigned facilities adjacent to each other with an aisle of 2 meters width between them. Thus $F_I$ is adjacent to $F_{II}$, $F_{III}$ is adjacent to $F_{IV}$ and $F_V$ is adjacent to $F_{VI}$.

Based on the cost information provided in the load-matrix the total cost in terms of load-units for the above layout can be calculated as follows:

$$L = 2\{(20 + 10) + (50 + 30) + (10 + 15)\} + (44 \times 25) + (22 \times 40) + (22 \times 15) = 2580.$$

By feeding the above optimal solution into the CRAFT program the final, the global optimum is found in a single iteration. The final, optimal layout as obtained by CRAFT is as under:

**Figure 3**

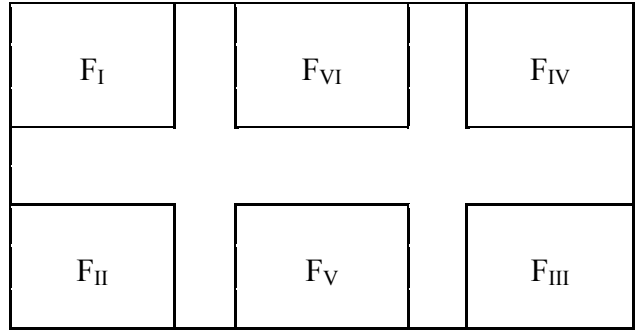

Based on the cost information provided in the load-matrix the total cost in terms of load-units for the optimal layout can be calculated as follows:

L* = 2{(10 + 20) + (15 + 10) + (5 + 30)} + (22 x 25) + (44 x 15) + (22 x 40) = 2360.

Therefore the final solution is an improvement of just 220 load-units over the initial solution! This shows that this initial solution fed into CRAFT is indeed near optimal and can thus ensure a faster convergence.

**References**

[1] Buffa, Elwood S., Armour G. C. and Vollmann, T. E. (1964), "Allocating Facilities with CRAFT", Harvard Business Review, Vol. 42, No.2, pp.136-158

[2] Carrie, A. S. (1980), "Computer-Aided Layout Planning – The Way Ahead", International Journal of Production Research, Vo. 18, No. 3, pp. 283-294

[3] Driscoll, J. and Sangi, N. A. (1988), "An International Survey of Computer-aided Facilities Layout – The Development And Application Of Software", Published

# The Israel-Palestine Question – A Case for Application of Neutrosophic Game Theory

Dr. Sukanto Bhattacharya

Business Administration Department

Alaska Pacific University

4101 University Drive

Anchorage, AK 99508, USA

Dr. Florentin Smarandache

Department of Mathematics and Statistics

University of New Mexico, U.S.A.

Dr. Mohammad Khoshnevisan

School of Accounting and Finance

Griffith University, Australia

**Abstract**

In our present paper, we have explored the possibilities and developed arguments for an application of principles of *neutrosophic game theory* as a generalization of the fuzzy game theory model to a better understanding of the Israel-Palestine problem in terms of the goals and governing strategies of either side. We build on an earlier attempted justification of a game theoretic explanation of this problem by Yakir Plessner (2001) and go on to argue in favour of a neutrosophic adaptation of the standard 2x2 zero-sum game theoretic model in order to identify an optimal outcome.

**Key Words:** Israel-Palestine conflict, Oslo Agreement, fuzzy games, neutrosophic semantic space

**Background**

There have been quite a few academic exercises to model the ongoing Israel-Palestine crisis using principles of statistical game theory. However, though the optimal solution is ideally sought in the identification of a *Nash equilibrium* in a cooperative game, the true picture is closer to a zero-sum game rather than a cooperative one. In fact it is not even a zero-sum game at all times, as increasing levels of mutual animosity in the minds of the players often pushes it closer to a sub-zero sum game. (Plessner, 2001).

As was rightly pointed out by Plessner (2001), the application of game theory methodology to the current conflict between Israel and the Palestinians is based on identifying the options that each party has, and an attempt to evaluate, based on the chosen option, what each of them is trying to achieve. The Oslo Agreement is used as an instance with PLO leadership being left to choose between two mutually exclusive options: either compliance with the agreement or non-compliance. Plessner contended that given the options available to PLO leadership as per the Oslo Agreement, the following are the five possible explanations for its conduct:

- The PLO leadership acts irrationally;

- Even though the PLO leadership wants peace and desires to comply, it is unable to do so because of mounting internal pressures;

- PLO leadership wants peace but is unwilling to pay the internal political price that any form of compliance shall entail;

- PLO leadership wants to keep the conflict going, and believes that Israel is so weak that it does not have to bear the internal political price of compliance, and can still achieve his objectives; or

- Given the fact that PLO leadership has been encouraging violence either overtly or covertly, it is merely trying to extract a better final agreement than the one achievable without violence

Plessner (2001) further argued that the main objective of the players is not limited to territorial concessions but rather concerns the recognition of Palestinian sovereignty over Temple Mount and the right of return of Palestinian refugees to pre-1967 Israel; within the territorial boundaries drawn at the time of the 1949 Armistice Agreements.

However, a typical complication in a problem of this kind is that neither the principal objective nor the strategy vectors remain temporally static. That is, the players' goals and strategies change over time making the payoff matrix a dynamic one. So, the same players under a similar set-up are sometimes found engaging in cooperative games and at other times in non-cooperative ones purely depending on their governing strategy vectors and principal objective at any particular point of time. For example, the PLO leadership may have bargained for a better final agreement using pressure tactics based on violence in the pre 9/11 scenario when the world had not yet woken up fully to the horrors of global terrorism and he perceived that the Israel was more likely to make territorial concessions in exchange of lasting peace. However, in the post 9/11 scenario, with the global opinion strongly united against any form of terrorism, its governing strategy vector will have to change as Israel now not only will stone-wall the pressure tactics, but will also enjoy more liberty to go on the offensive.

Moreover, besides being temporally unstable, the objectives and strategies are often ill-defined, inconsistent and have a lot of interpretational ambiguity. For example, while a strategy for the PLO leadership could appear to be keeping the conflict alive with the covert objective of maintaining its own organizational significance in the Arabian geopolitics, at the same time there would definitely have to be some actions from its side which would convey a clear message to the other side that it wants to end the conflict – which apparently has been its overt objective, which would then get Israel to reciprocate its overt intentions. But in doing so, Israel could gain an upper hand at the bargaining table, which would again cause internal pressures to mount on PLO leadership thereby jeopardizing the very position of power it is seen trying to preserve by keeping the conflict alive.

**The problem modelled as a standard 2x2 zero-sum game**

<div align="center">

**<u>Palestine</u>**

|  | I | II |
|---|---|---|
| **I** | 1 | -1 |
| **II** | 0 | -1 |
| **III** | 0 | -1 |
| **IV** | 1 | 0 |

</div>

(Row labels I, II, III, IV are for **<u>Israel</u>**.)

Palestine's strategy vector: (I – full compliance with Oslo Agreement, II – partial or non-compliance)

Israel's strategy vector: (I - make territorial concessions, II - accept right of return of the Palestinian refugees, III – launch an all-out military campaign, IV – continue stone-walling)

The payoff matrix has been constructed with reference to the row player i.e. Israel. In formulating the payoff matrix it is assumed that combination (I, I) will potentially end the conflict while combination (IV, II) will basically mean a status quo with continuing conflict. If Palestine can get Israel to either make territorial concessions or accept the right of return of Palestinian refugees without fully complying with the Oslo Agreement i.e. strategy combinations (I, II) and (II, II), then it marks a gain for the former and a loss for the latter. If Israel accepts the right of return of Palestinian refugees and Palestine agrees to fully comply with the Oslo Agreement, then though it would potentially end the conflict, it could possibly be putting the idea of an independent Jewish state into jeopardy and so the strategy combination (II, I) does not have a positive payoff for Israel. If Israel launches an all-out military campaign and forces Palestine into complying with the Oslo Agreement i.e. strategy combination (III, I) then it would not result in an exactly positive payoff for Israel due to possible alienation of world opinion and may be even losing some of the U. S. backing. If an all-out Israeli military aggression causes a hardening of stance by Palestine then it will definitely result in a negative payoff due to increased violence and bloodshed. If however, there is a sudden change of heart within the Palestinian leadership and Palestine chooses to fully abide by the Oslo Agreement without any significant corresponding territorial or political consideration by Israel i.e. strategy combination (IV, I), it will result in a potential end to the conflict with a positive payoff for Israel.

In the payoff matrix, the last row dominates the first three rows while the second column dominates the first column. Therefore the above game has a *saddle point* for the strategy combination (IV, II) which shows that in their attempt to out-bargain each other both parties will actually end up continuing the conflict indefinitely!

It is clear that Palestine on its part will not want to ever agree to have full compliance with the Oslo Agreement as it will see always see itself worse off that way. Given that Palestine will never actually comply fully with the Oslo Agreement, Israel will see in its best interest to continue the status quo with an ongoing conflict, as it will see itself ending up on the worse end of the bargain if it chooses to play any other strategy.

The equilibrium solution as we have obtained here is more or less in concurrence with the conclusion reached by Plessner. He argued that given the existing information at Israel's disposal, it is impossible to tell whether PLO leadership chooses non-compliance because it will have to pay a high internal political price otherwise or because it may want to keep the conflict alive just to wear down the other side thereby opening up the possibility of securing greater bargaining power at the negotiating table. The point Plessner sought to make is that whether or not PLO leadership truly wants peace is immaterial because in any case it will act in order to postpone a final agreement, increase its weight in the international political arena and also try to gain further concessions from Israel.

**Case for applying neutrosophic game theory**

However, as is quite evident, none of the strategy vectors available to either side will remain temporally stationary as crucial events keep unfolding on the global political stage in general and the Middle-Eastern political stage in particular. Moreover, there is a lot of ambiguity about the driving motives behind PLO leadership's primary goal and the strategies it adopts to achieve that goal. Also it is hard to tell apart a true bargaining strategy from one just meant to be a political decoy. This is where we believe and advocate an application of the conceptual framework of the neutrosophic game theory as a generalization of the dynamic fuzzy game paradigm.

In generalized terms, a well-specified dynamic game at time t is a particular interaction ensemble with well defined rules and roles for the players within the ensemble, which remain in place at time t but are allowed to change over time. However, the players often may suffer from what is termed in organizational psychology as "role ambiguity" i.e. a situation where none of the players are exactly sure what to expect from the others or what the other players expect from them. In the context of the Israel-Palestine problem, for example, PLO leadership would probably not have been sure of its exact role when Yasser Arafat met with U.S. and Israeli leadership at the Camp David Summit ostensibly to hammer out a peace agreement. Again following Plessner's argument, Arafat went to that Summit against his free will and would have liked to avoid Camp David if he could because he did not want to sign any final agreement that was short of a complete renunciation of its sovereign existence by Israel. With no such capitulation forthcoming from Israel, it was in PLO leadership's best interest to keep the conflict alive. However, it did have to give certain overt indications mainly to keep U.S. satisfied that a negotiated settlement was possible and was being preferred over letting loose Hamas mercenaries on the streets. Under such circumstances, it would be quite impossible to pick out a distinct governing strategy which the other side could then meet with a counter-strategy.

However, one positive aspect about Summits such as the Camp David Summit is that they make the game scenario an open one in the sense that the conflicting parties are able to dynamically construct and formulate objectives and strategies in the course of their peaceful, mutual interaction within a formally defined socio-political set-up. This allows a closer analytical study of the negotiation process where the negotiation space may be defined as $N_{Palestine} \cap N_{Israel}$.

There is a *fuzzy semantic space* which is a collective of each player's personal views about what constitutes a "just deal" (Burns and Rowzkowska, 2002). Such views are formed based on personal value judgments, past experience and also an expectation about the possible best-case and worst-case negotiation outcomes. This fuzzy semantic space is open to modifications as negotiations progress and views are exchanged resulting in earlier notions being updated in the light of new information.

This semantic space however remains fuzzy due to vagueness about the exact objectives and lack of precise understanding of the exact stakes which the opposing parties have *from their viewpoints*. That is to say, none of the conflicting parties can effectively put themselves in the shoes of each other and precisely understand each other's nature of expectations.

This is borne out in the Camp David Summit when probably one side of the table was thinking in terms of keeping the conflict alive while giving the impression to the other side that they were seriously seeking ways to end it. This immediately makes it clear why such a negotiation would break down, simply because it never got started in the first place!

If the Israel-Palestine problem is formulated as a dynamic fuzzy bargaining game, the players' *fuzzy set judgement functions* over expected outcomes may be formulated as follows according to the well known rules of fuzzy set algebra (Zadeh, 1965):

For Palestine, the fuzzy evaluative judgment function at time t, **J (P, t)** will be given by the fuzzy set membership function $M_{J(P, t)}$ which is expressed as follows:

$$M_{J(P, t)}(x) = \begin{cases} c \in (0.5, 1); & \text{for } \wp_{Worst} < x < \wp_{Best} \\ 0.5; & \text{for } x = \wp_{Worst}; \text{ and} \\ 0; & \text{for } x \leq \wp_{Worst} \end{cases}$$

Here $\wp_{Best}$ is the best possible negotiation outcome Palestine could expect; which, according to Plessner, would probably be Israeli recognition of the right of return of Palestinian refugees to their pre-1967 domicile status. For Israel on the other hand, the fuzzy evaluative judgment function at time t, **J (I, t)** will be given by the fuzzy set membership function $M_{J(I, t)}$ which will be as follows:

$$M_{J(I, t)}(y) = \begin{cases} 1; & \text{for } y \geq \Im_{Best} \\ c' \in (0.5, 1); & \text{for } \Im_{Worst} < y < \Im_{Best} \\ 0.5; & \text{for } y = \Im_{Worst}; \\ 0; & \text{for } y \leq \Im_{Worst} \end{cases}$$

Here $\Im_{Worst}$ is the worst possible negotiation outcome Israel could expect; which, would concur with the best expected outcome for Palestine.

However, the semantic space $N_{Palestine} \cap N_{Israel}$ is more generally framed as a *neutrosophic semantic space* which is a three-way generalization of the fuzzy semantic space and includes a third, neutral possibility whereby the semantic space cannot be de-fuzzified into two crisp zero-one states due to the incorporation of an intervening state of "indeterminacy". Such indeterminacy could practically arise from the fact that any mediated, two-way negotiation process is likely to become *over-catalyzed* by the subjective utility preferences of the mediator – in case of the Israel-Palestine problem; that of the U.S. (and to a lesser extent; that of some of the other permanent members of the UN Security Council).

*Neutrosophy* is a new branch of philosophy that is concerned with *neutralities* and their interaction with various ideational spectra (Smarandache, 2000). Let *T, I, F* be real subsets of the *non-standard interval* $]^-0, 1^+[$. If $\epsilon > 0$ is an infinitesimal such that for all positive integers n and we have $|\epsilon| < 1/n$, then the non-standard finite numbers $1^+ = 1+\epsilon$ and $0^- = 0-\epsilon$ form the boundaries of the non-standard interval $]^-0, 1^+[$. Statically, *T, I, F* are *subsets* while dynamically, as in our case when we are using the model in the context of a dynamic game, they may be viewed as *set-valued vector functions*. If a logical proposition is said to be *t%* true in *T*, *i%* indeterminate in *I* and *f%* false in *F* then *T, I, F* are referred to as the *neutrosophic components*. Neutrosophic probability is useful to events that are shrouded in a *veil of indeterminacy* like the *actual* implied volatility of long-term options. As this approach uses a *subset-approximation* for truth values, indeterminacy and falsity-values it provides a better approximation than classical probability to uncertain events.

Therefore, for Palestine, the neutrosophic evaluative judgment function at time t, $\mathbf{J_N (P, t)}$ will be given by the *neutrosophic set membership function* $M_{JN (P, t)}$ which may be expressed as follows:

$$M_{JN (P, t)} (x) = \begin{cases} c \in (0.5, 1); & \text{for } \wp_{Worst} < x < \wp_{Best} \textbf{ AND } x \in T \\ 0.5; & \text{for } x = \wp_{Worst} \textbf{ AND } x \in T \\ 0; & \text{for } x \leq \wp_{Worst} \textbf{ AND } x \in T \end{cases}$$

For Israel on the other hand, the neutrosophic evaluative judgment function at time t, $\mathbf{J_N (I, t)}$ will be given by the *neutrosophic set membership function* $M_{JN (I, t)}$ which may be expressed as follows:

$$M_{JN(I,t)}(y) = \begin{cases} 1; & \text{for } y \geq \Im_{Best} \text{ AND } y \in F \\ c' \in (0.5, 1); & \text{for } \Im_{Worst} < y < \Im_{Best} \text{ AND } y \in F \\ 0.5; & \text{for } y = \Im_{Worst} \text{ AND } y \in F; \\ 0; & \text{for } y \leq \Im_{Worst} \text{ AND } y \in F \end{cases}$$

Pertaining to the three-way classification of neutrosophic semantic space, it is $t$% true in sub-space $T$ that a mediated, bilateral negotiation will produce a favorable outcome within the evaluative judgment space of the Palestinian leadership while it is $f$% false in sub-space $F$ that the outcome will be favorable within the evaluative judgment space of the Palestinian leadership. However there is an $i$% indeterminacy in sub-space $I$ whereby the nature of the outcome may be neither favorable nor unfavorable within the evaluative judgment space of either competitor – for example if the negotiation process is over-catalyzed by the utility preferences of the mediator!

**Conclusion**

$M_{JN(P,t)}(x)$ {or $M_{JN(I,t)}(y)$} would be interpreted as Palestine's (or Israel's) degree of satisfaction with the negotiated settlement. Following Plessner's argument again, it is PLO leadership's ultimate objective to see the end of an independent Jewish state of Israel and if that be the case then of course there will always be an unbridgeable incongruence between $M_{JN(P,t)}(x)$ and $M_{JN(I,t)}(y)$ because of mutually inconsistent evaluative judgment spaces between the two parties to the conflict. Therefore, for any form of negotiation to have a positive result the first and foremost requirement would be to make the evaluative judgment spaces consistent. Because unless the evaluative judgment spaces are consistent, the negotiation space will degenerate into a null set at the very onset of the bargaining process thereby pre-empting any equilibrium solution different from the status quo. However, by its very definition, since these spaces are not crisp, they are malleable to some extent (Reiter, 1980). That is, even while retaining their core forms, subtle changes could be induced to make these spaces workably consistent. Then the aim of the mediator should to make the parties redefine their primary objectives without necessarily feeling that such redefinition itself means a concession. When this required redefinition of primary objectives has been achieved can the evaluative judgment spaces generate a negotiation space that will not become null *ab initio*. However, there is also an indeterminate aspect to any process of mediated bilateral dialogues between the two parties due to the catalyzation effect brought about by the subjective utility preferences of the mediator (or mediators).

# Effective Number of Parties in A Multi-Party Democracy Under an Entropic Political Equilibrium with Floating Voters

Sukanto Bhattacharya

Department of Business Administration

Alaska Pacific University, AK 99508, USA

Florentin Smarandache

University of New Mexico

200 College Road, Gallup, USA

**Abstract**

In this short, technical paper we have sought to derive, under a posited formal model of political equilibrium, an expression for the effective number of political parties (ENP) that can contest elections in a multi-party democracy having a plurality voting system(also known as a first-past-the-post voting system). We have postulated a formal definition of political equilibrium borrowed from the financial market equilibrium whereby given the set of utility preferences of all eligible voters as well as of all the candidates, each and every candidate in an electoral fray stands the same objective chance of getting elected. Using an expected information paradigm, we show that under a condition of political equilibrium, the effective number of political parties is given by the reciprocal of the proportion of core electorate (non-floating voters). We have further argued that the formulated index agrees with a party system predicted by Duverger's law.

**Key words**: Plurality voting, entropic equilibrium, floating voters, Duverger's law

**Introduction**

Plurality voting systems are currently used in over forty countries worldwide which include some of the largest democracies like USA, Canada, India and UK. Under the basic plurality voting system, a country is divided into territorial single-member constituencies; voters within each constituency cast a single ballot (typically marked by a X) for one candidate; and the candidate with the largest share of votes in each seat is returned to office; and the political party (or a confederation of ideologically similar political parties) with an overall majority of seats forms the government. The fundamental feature of the plurality voting system is that single-member constituencies are based on the size of the electorate. For example, the US is divided into 435 Congressional districts each including roughly equal populations with one House representative per district. Boundaries of constituencies are reviewed at periodic intervals based on the national census to maintain the electorate balance. However the number of voters per constituency varies dramatically across countries e.g. India has 545 representatives for a population of over nine hundred million, so each member of the Lok Sabha (House of the People) serves nearly two million people, while in contrast Ireland has 166 members in the Dial for a population slightly more than three-and-half million or approximately one seat for a little over twenty thousand people.

Under the first-past-the-post voting system candidates only need a simple plurality i.e. at least one more vote than their closest rival to get elected. Hence in three-way electoral contests, the winning candidate can theoretically have less than fifty percent of votes cast in his or her favor e.g. if the vote shares are 35%, 34% and 31%, the candidate with a 35% vote share will get elected. Therefore, although two-thirds of voters support other candidates, the candidate with a simple plurality of votes wins the contest (Norris, 1997).

We define political equilibrium as a condition in which the choices of voters and political parties are all compatible and in which no one group can improve its position by making a different choice. In essence therefore, political equilibrium may be said to exist when, given the set of utility preferences of all eligible voters as well as of all the candidates, each and every candidate in an electoral fray stands the same chance of getting elected. This definition is adequately broad to cover more specific conditional equilibrium models and is based on the principle of efficiency as applied to financial markets. Daniel Sutter (2002) defines political equilibrium as "a balance between demands by citizens on the political system and candidates compete for office". Therefore, translated to a multi-party democracy having a plurality voting system, political equilibrium can be thought to imply a state where perfect balance of power exists between all contesting parties. Methodologically, we build our formal equilibrium model using an expected information approach used in a generalized financial market equilibrium model (Bhattacharya, 2001).

**Computing an effective number of political parties**

Is there a unique optimum for the number of political parties that have to compete in order to ensure a political equilibrium? If there indeed is such an optimal number then this number necessarily has to be central to any theoretical formalization of political equilibrium as we have defined. Rae (1967) advanced the first formal expression for political fractionalization in a multi-party democracy as follows:

$$F_s = 1 - \sum (s_i)^2$$

Here $F_s$ is known as Rae's index of political fractionalization and $s_i$ is the proportion of seats of the $i^{th}$ political party in the Parliament. Conceptually, Rae's fractionalization index is adapted from the Herfindahl-Hirschman market power concentration index. F is 0 for a single-party system and F tends to 0.50 for a two-party system in equilibrium i.e. when both parties command same proportion of seats in the Parliament. Of course F asymptotically approaches unity as the party system becomes more and more fractionalized. Of course, one may adapt Rae's fractionalization index in terms of the proportion of votes secured in an election instead of seats in Parliament. In that case Rae's index of fractionalization may be represented as follows:

$$F_v = 1 - \sum (v_i)^2$$

Dumont and Caulier (2003) have recognized two major drawbacks of Rae's index. Firstly, the index is not linear for parties that are tied in strength; measured either as proportion of seats or proportion of votes. A two-party system in equilibrium produces an F of 0.50 whereas a four-party system in equilibrium produces 0.75 and a five-party system in equilibrium will have an F of 0.80. Dumont and Caulier (2003) point out that this feature makes the F untenable as an index as the operationalized measure and the phenomenon it measures follow different progression paths. Secondly, Rae's index is, like most other normalized indices of social phenomena, extremely difficult to interpret in objective terms as a unique variable characterizing a party system. The effective number of parties (ENP) measure formulated by Laakso and Taagepera (1979) by improving on Rae's index is now commonly regarded as the classical numerical measure for the comparative analysis of party systems. This ENP formula takes both the number of parties and their relative weights into account when computing a unique variable characterizing a party system thereby making objective interpretation a lot easier as compared to Rae's fractionalization index. The ENP formula is simply stated as the reciprocal of the complement of Rae's fractionalization index i.e.

$$ENP_s = (1 - F_s)^{-1} \text{ and } ENP_v = (1 - F_v)^{-1}$$

In equilibrium, all political parties will command the same strength measured either as proportion of seats or votes and ENP will exactly equal the number of parties in fray. Taagepera and Shugart (1989) have argued that the ENP has become a widely-used index because it "usually tends to agree with our average intuition about the number of serious parties". However Molinar (1991) and Dunleavy and Boucek (2003) have argued that this index produces counter-intuitive and counter-empirical results under a number of circumstances. Taagepera (1999) himself suggested that in cases where one party clearly dominates the political system (commanding more than 50% of the seats), an additional index called the LC (Largest Component) index should be used in conjunction with ENP. The LC is simply the reciprocal of the share of the largest party. When LC is greater than 2 for any party, that party clearly dominates the political system which would however be classified as a multi-party system if only the ENP was the sole classification criterion. Dunleavy and Boucek (2003) have advocated the averaging of ENP index with the LC index to yield a unique classification criterion. Dumont and Caulier (2003) advanced the effective number of relevant parties measure (ENRP) as an improvement over the ENP in a way that their measure yields a unique classification criterion that roughly corresponds to the ENP measure when there are more than two parties that can be considered as major contenders for victory in an electoral contest and collapses to unity if there are only one or two parties that can be seriously considered as a potential winner.

Irrespective of which variant of the ENP index we consider, it is obvious that an intuitive paradigm formalizing political equilibrium in a multi-party democracy having a plurality voting system may be constructed if it can be shown that in equilibrium, all parties in fray are indeed expected to command an equal strength measured either in terms of seats or votes. But such formalization would be considered somewhat limited if it did not take into account the impact of floating voters on electoral outcomes. These are the quintessential fence-sitters who waver between parties during the course of a Parliament, or who don't make up their minds until very close to the election (or even until actually putting their stamps on the ballot paper). The impact of floating voters on electoral outcome is all the more an important issue for large-sized electorates as is the case for very populous countries like India. But none of the ENP indices consider floating voters.

Effective number of political parties with floating voters in entropic equilibrium
Considering a finite fraction of floating voters in any electorate, we may define the following relationship as the (conservative) expected vote share of the $i^{th}$ political party:

$$E(V_i) = [E(S_i)](1 - \lambda_i)$$

Here $E(S_i)$ is the $i^{th}$ candidate's expected vote share as a proportion of the total electorate size and $\lambda_i$ is the fraction of the $i^{th}$ candidate's vote share that is deemed to come from floating voters. This is the fraction of electorate which is generally supportive of the $i^{th}$ candidate but this support may or may not be translated into actual votes on the day of the election. Thus $E(S_i)$ is the expected proportion of votes to be cast in the $i^{th}$ candidate's favor accepting the existence of floating voters in the electorate. Therefore we may write:

$$\sum_i E(V_i) = \sum_i [E(S_i)](1 - \lambda_i)$$

Let us denote $\sum_i E(V_i)$ as $E(V)$ and $\sum_i E(S_i)$ as $E(S)$. Therefore, re-arranging (5) we get:

$$\sum_i [E(S_i)] \lambda_i = E(S) - E(V)$$

In mathematical information theory, entropy or expected information from an event is measured using a logarithmic function borrowed from classical thermodynamics. There are two possible mutually exclusive and exhaustive outcomes for any individual event – either the event occurs or the event does not occur. If there are m candidates in an electoral fray the two events associated with each candidate in fray is that either the particular candidate wins the election or he/she does not win. If $p_i$ is the probability of the $i^{th}$ candidate winning the election, then the expected information content of a message that conveys the outcome of an election with i = 1, 2, …, m candidates is obtained by the classical entropy function as formulated by Shannon (1948) as follows:

$$\psi(p) = (-C')\sum_i (p_i)\log_2(p_i)$$

Here $C'$ is a positive scale factor (a negentropic counterpart of the Boltzmann constant in thermodynamic entropy). Under an m-party political equilibrium, the long run core (non-floating) vote shares of the i = 1, 2, …, m candidates in electoral fray may be considered as equivalent to their long run winning probabilities. Thus $\psi(p)$ is re-writable as follows:

$$\psi(1 - \lambda) = (-C')\sum_i (1 - \lambda_i)\log_2(1 - \lambda_i)$$

**Proposition:** If $\psi(1 - \lambda)$ is the expected information from the knowledge of an electoral outcome given the proportion of non-floating voters $(1 - \lambda_i)$ in the vote share of the $i^{th}$ candidate, then the effective number of parties under entropic equilibrium is given as:

$$ENP(\lambda) = (1 - \lambda^*)^{-1}; \text{ where } \lambda^* = 1 - E(V)/E(S)$$

**Proof:** Incorporating the Lagrangian multiplier L the objective function can be written as:

$$Z(1 - \lambda_i, L) = (-C') \sum_i (1 - \lambda_i) \log_2 (1 - \lambda_i) + L\{1 - \sum_i (1 - \lambda_i)\}$$

Taking partial derivative of Z with respect to $(1 - \lambda_i)$ and setting equal to zero as per the necessary condition of maximization, the following stationary condition is obtained:

$$\partial Z / \partial (1 - \lambda_i) = (-C')\{\log_2(1 - \lambda_i) + 1\} - L = 0$$

Therefore at the point of maximum entropy one gets $\log_2(1 - \lambda_i) = -(L/C' + 1)$ i.e. $(1 \lambda_i)$ becomes a constant value independent of i for all i = 1, 2, …, m candidates in the electoral contest. Since necessarily the $1 - \lambda_i$ values must sum to unity, it implies that at the point of maximum entropy we must have $p_1 = p_2 = \ldots = p_m = (1 - \lambda^*) = 1/m$.

$$\text{Therefore } m \equiv ENP(\lambda) = (1 - \lambda^*)^{-1}$$

Simplifying the expression for $\sum_i [E(S_i)] \lambda_i = E(S) - E(V)$ under equilibrium we may write:

$$\lambda^* E(S) = E(S) - E(V) \text{ i.e. } \lambda^* = 1 - E(V)/E(S) \qquad \text{Q.E.D.}$$

$\lambda^*$ is simply the total percentage of floating voters under an entropic political equilibrium[1]. Thus $ENP(\lambda)$ is formally obtained (as expected intuitively) as the reciprocal of the equilibrium percentage of non-floating voters in the electorate. The higher the proportion of floating voters within the electorate, the higher is the value of $ENP(\lambda)$. The intuitive reasoning is obvious – with a large number of floating votes to go around, more candidates could stay in the electoral fray than there would be if the electorate consisted of only a very small percentage of floating voters. When $\lambda = 50\%$, $ENP(\lambda) = 2$. If $\lambda$ goes up to 75%, $ENP(\lambda)$ will go up to 4 i.e. with 25% more floating voters within the electorate, 2 more candidates can stay in electoral fray feeding off the floating votes.

Thus $ENP(\lambda)$ (the formula for which is structurally quite similar to Laakso and Taagepera's ENP index) is a generalized measure of ENP based on the entropic formalization of political equilibrium accepting the very real existence of floating voters.

**Entropic political equilibrium and Duverger's law**

Duverger (1951) stated that the electoral contest in a single-seat electoral constituency following a plurality voting system tends to converge to a two-party system. *Duverger's law* basically stems from the premise of *strategic voting*. Palfrey (1989) has showed that in large electorates, equilibrium voting behavior implies that a voter will always vote for the most preferred candidate of the two frontrunners. For a given electorate of size n, Palfrey's model is stated in terms of the following inequality:

$$u_k > u_j \left[ \left(\sum_{i \neq j}(p^n_{ij}/p^n_{kl})\right) / \left(\sum_{h \neq k}(p^n_{kh}/p^n_{kl})\right) \right] + \sum_{i \neq j,k} u_i \left[ \{(p^n_{ki} - p^n_{ij})/p^n_{kl}\} / \sum_{h \neq k}(p^n_{kh}/p^n_{kl}) \right]$$

In this model, $u_k$ denotes the voter's utility of his/her first choice among the two frontrunners and $u_l$ denotes the voter's utility for his/her second choice among the frontrunners so that $u_k > u_l$. Also j is any other candidate from among the i = 1, 2, ..., m candidates. The notation $p^n_{ij}$ stands for the probability that the candidate i and candidate j are tied for the most votes and the interpretation is similar for notations $p^n_{kh}$ and $p^{nkl}$. In the limiting case, the likelihood ratio $p^n_{kh}/p^{nkl}$ tends to zero for all $ij \neq kl$. Thus the right-hand side of the inequality converges to $u_l$ irrespective of j; thereby mathematically establishing Duverger's law. Apart from Palfrey's theoretical formalization, Cox and Amorem Neto (1997) and Benoit (1998) and Schneider (2004) have provided empirical evidence generally supportive of Duverger's law.

It therefore seems rather appropriate that an intuitive model of political equilibrium in a multi-party democracy that follows a plurality voting system should at least take Duverger's law into consideration if not actually have it embedded in some form within its formal structure. This is true for our entropic model, because as m increases $(1 - \lambda^*) = {}^1/_m$ becomes smaller and smaller, thereby implying that for multi-party democracies that follow a plurality voting system, the political equilibrium most likely to prevail in the long run will tend to occur at the highest possible value of $(1 - \lambda^*) = 50\%$. In other words, although some relatively new democracies may start off with a number of political parties contesting elections and a very large percentage of floating voters in the electorate, the likelihood is very low that a very high proportion (exceeding 50%) of the electorate will be composed of floating voters in the long run which implies that in the long run, "mature" multi-party democracies having plurality voting systems will tend to have only two parties as serious contenders for victory in an election; corresponding to a two-party system as stated by Duverger's law.

**Conclusion**

We have proposed and mathematically derived a formula for the effective number of political parties that can be in electoral fray under a condition of political equilibrium in a multi-party democracy following a plurality voting system. We have posited the expected information approach to formalize the concept of political equilibrium in a parliamentary democracy. Our advocated model aims to improve upon existing ENP indices by incorporating the very realistic consideration of the impact of floating voters on elections. Of course, ours has been an entirely theoretical exercise and a potentially rewarding direction of future research would be to empirically investigate the veracity of ENP(λ) possibly in conjunction with a suitable classification model to distinguish floating voters.

# Notion of Neutrosophic Risk and Financial Markets Prediction

Dr. Sukanto Bhattacharya

Program Director - MBA Global Finance

Business Administration Department

Alaska Pacific University

4101 University Drive

Anchorage, AK 99508, USA

**Abstract**

In this paper we present an application of the neutrosophic logic in the prediction of the financial markets.

## 1. Introduction

The *efficient market hypothesis* based primarily on the statistical principle of *Bayesian inference* has been proved to be only a special-case scenario. The generalized financial market, modeled as a *binary, stochastic system* capable of attaining one of two possible states (High → 1, Low → 0) with finite probabilities, is shown to reach *efficient equilibrium* with **p . M = p** if and only if the transition probability matrix $\mathbf{M_{2x2}}$ obeys the additionally imposed condition $\{m_{11} = m_{22}, m_{12} = m_{21}\}$, where $m_{ij}$ is an element of M (Bhattacharya, 2001). [1]

Efficient equilibrium is defined as the stationery condition **p** = [0.50, 0.50] i.e. the state in t + 1 is equi-probable between the two possible states given the market vector in time t. However, if this restriction $\{m_{11} = m_{22}, m_{12} = m_{21}\}$ is removed, we get inefficient *equilibrium* **ρ** = $[m_{21}/(1-v), m_{12}/(1-v)]$, where $v = m_{11} - m_{21}$ may be derived as the *eigenvalue of* **M** and **ρ** is a generalized version of **p** whereby the elements of the market vector are no longer restricted to their efficient equilibrium values. Though this proves that the generalized financial market cannot possibly get reduced to *pure random walk* if we do away with the *assumption of normality*, it does not necessarily rule out the possibility of *mean reversion* as M itself undergoes transition over time implying a probable re-establishment of the condition $\{m_{11} = m_{22}, m_{12} = m_{21}\}$ at some point of time in the foreseeable future. The temporal drift rate may be viewed as the *mean reversion parameter* k such that $\mathbf{k^j M_t \rightarrow M_{t+j}}$. In particular, the options market demonstrates a rather perplexing departure from efficiency. In a *Black-Scholes type world*, if stock price volatility is known *a priori*, the option prices are completely determined and any deviations are quickly arbitraged away.

Therefore, statistically significant mispricings in the options market are somewhat unique as the only non-deterministic variable in option pricing theory is volatility. Moreover, given the knowledge of implied volatility on the short-term options, the miscalibration in implied volatility on the longer term options seem odd as the parameters of the process driving volatility over time can simply be estimated by an AR1 model (Stein, 1993). [2]

Clearly, the process is not quite as straightforward as a simple parameter estimation routine from an autoregressive process. Something does seem to affect the market players' collective pricing of longer term options, which clearly overshadows the straightforward considerations of implied volatility on the short-term options. One clear reason for inefficiencies to exist is through *overreaction* of the market players to new information. Some inefficiency however may also be attributed to purely *random white noise* unrelated to any coherent market information. If the process driving volatility is indeed mean reverting then a low implied volatility on an option with a shorter time to expiration will be indicative of a higher implied volatility on an option with a longer time to expiration. Again, a high implied volatility on an option with a shorter time to expiration will be indicative of a lower implied volatility on an option with a longer time to expiration. However statistical evidence often contradicts this *rational expectations hypothesis* for the *implied volatility term structure*.

Denoted by $\sigma'_t(t)$, (where the symbol ' indicates first derivative) the implied volatility at time t of an option expiring at time T is given in a Black-Scholes type world as follows:

$$\sigma'_t(t) = {}_{j=0}\int^T [\{\sigma_M + k^j (\sigma_t - \sigma_M)\}/T]\, dj$$

$$\sigma'_t(t) = \sigma_M + (k^T - 1)(\sigma_t - \sigma_M)/(T \ln k) \qquad (1)$$

Here $\sigma_t$ evolves according to a *continuous-time, first-order Wiener process* as follows:

$$d\sigma_t = -\beta_0 (\sigma_t - \sigma_M)\, dt + \beta_1 \sigma_t\, \varepsilon \sqrt{dt} \qquad (2)$$

$\beta_0 = -\ln k$, where k is the *mean reversion parameter*. Viewing this as a *mean reverting AR1 process* yields the expectation at time t, $E_t(\sigma_{t+j})$, of the instantaneous volatility at time t+j, in the required form as it appears under the integral sign in equation (1).

This theorizes that volatility is rationally expected to gravitate geometrically back towards its long-term mean level of $\sigma_M$. That is, when instantaneous volatility is above its mean level ($\sigma_t > \sigma_M$), the implied volatility on an option should be decreasing as $t \to T$. Again, when instantaneous volatility is below the long-term mean, it should be rationally expected to be increasing as $t \to T$. That this theorization does not satisfactorily reflect reality is attributable to some kind combined effect of overreaction of the market players to *excursions in implied volatility of short-term options* and their *corresponding underreaction to the historical propensity of these excursions to be rather short-lived*.

## 2. A Cognitive Dissonance Model of Behavioral Market Dynamics

Whenever a group of people starts acting in unison guided by their hearts rather than their heads, two things are seen to happen. Their individual suggestibilities decrease rapidly while the suggestibility of the group as a whole increases even more rapidly. The 'leader', who may be no more than just the most vociferous agitator, then primarily shapes the groupthink. He ultimately becomes the focus of the group opinion. In any financial market, it is the gurus and the experts who often play this role. The crowd hangs on their every word and makes them the uncontested Oracles of the marketplace.

If figures and formulae continue to speak against the prevailing groupthink, this could result into a *mass cognitive dissonance* calling for reinforcing self-rationalizations to be strenuously developed to suppress this dissonance. As individual suggestibilities are at a lower level compared to the group suggestibility, these self-rationalizations can actually further fuel the prevailing groupthink. This groupthink can even crystallize into something stronger if there is also a simultaneous *vigilance depression effect* caused by a tendency to filter out the dissonance-causing information. The non-linear feedback process keeps blowing up the bubble until a critical point is reached and the bubble bursts ending the prevailing groupthink with a recalibration of the position by the experts.

Our proposed model has two distinct components – a *linear feedback process* containing no looping and a *non-linear feedback process* fuelled by an *unstable rationalization loop*. It is due to this loop that perceived true value of an option might be pushed away from its theoretical true value. The market price of an option will follow its *perceived true value* rather than its *theoretical true value* and hence the inefficiencies arise. This does not mean that the market as a whole has to be inefficient – the market can very well be close to strong efficiency! Only it is the perceived true value that determines the actual price-path meaning that all market information (as well as some of the random white noise) gets automatically *anchored* to this perceived true value. This would also explain why excursions in short-term implied volatilities tend to dominate the historical considerations of mean reversion – the perceived term structure simply becomes anchored to the prevailing groupthink about the nature of the implied volatility.

Our conceptual model is based on two primary assumptions:

The *unstable rationalization loop* comes into effect if and only if the group is a reasonably well-bonded one i.e. if the initial group suggestibility has already attained a certain minimum level as, for example, in cases of strong cartel formations and;

The *unstable rationalization loop* stays in force till some critical point in time t* is reached in the life of the option. Obviously t* will tend to be quite close to T – the time of expiration. At that critical point any further divergence becomes unsustainable due to the extreme pressure exerted by real economic forces 'gone out of sync' and the gap between perceived and theoretical true values close very rapidly.

## 2.1. The Classical Cognitive Dissonance Paradigm

Since Leon Festinger presented it well over four decades ago, cognitive dissonance theory has continued to generate a lot of interest as well as controversy. [3] [4] This was mainly due to the fact that the theory was originally stated in very generalized, abstract terms. As a consequence, it presented possible areas of application covering a number of psychological issues involving the interaction of cognitive, motivational, and emotional factors. Festinger's dissonance theory began by postulating that pairs of cognitions (elements of knowledge), given that they are relevant to one another, can either be in agreement with each other or otherwise. If they are in agreement they are said to be *consonant*, otherwise they are termed *dissonant*. The mental condition that forms out of a pair of dissonant cognitions is what Festinger calls *cognitive dissonance*.

The existence of dissonance, being psychologically uncomfortable, motivates the person to reduce the dissonance by a process of *filtering out* information that are likely to increase the dissonance. The greater the degree of the dissonance, the greater is the pressure to reduce dissonance and change a particular cognition. The likelihood that a particular cognition will change is determined by the *resistance to change* of the cognition. Again, resistance to change is based on the *responsiveness* of the cognition to reality and on the extent to which the particular cognition is in line with various other cognitions. Resistance to change of cognition depends on the extent of loss or suffering that must be endured and the satisfaction or pleasure obtained from the behavior. [5] [6] [7] [8] [9] [10] [11] [12]

We propose the conjecture that cognitive dissonance is one possible (indeed highly likely) *critical behavioral trigger* [13] that sets off the rationalization loop and subsequently feeds it.

## 2.2 Non-linear Feedback Statistics Generating a Rationalization Loop

In a linear autoregressive model of order R, a time series $y_n$ is modeled as a linear combination of N earlier values in the time series, with an added correction term $x_n$:

$$y_n = x_n - \Sigma a_j\, y_{n-j} \qquad (3)$$

The autoregressive coefficients $a_j$ ($j = 1, \ldots N$) are fitted by minimizing the mean-squared difference between the modeled time series $y_n$ and the observed time series $y_n$. The minimization process results in a system of linear equations for the coefficients $a_n$, known as the ***Yule-Walker equations***. Conceptually, the time series $y_n$ is considered to be the output of a discrete linear feedback circuit driven by a noise $x_n$, in which delay loops of lag j have *feedback strength* $a_j$. For Gaussian signals, an autoregressive model often provides a concise description of the time series $y_n$, and calculation of the coefficients $a_j$ provides an indirect but highly efficient method of spectral estimation. In a full nonlinear autoregressive model, quadratic (or higher-order) terms are added to the linear autoregressive model. A constant term is also added, to counteract any net offset due to the quadratic terms:

$$y_n = x_n - a_0 - \Sigma a_j\, y_{n-j} - \Sigma b_{j,k}\, y_{n-j} y_{n-k} \qquad (4)$$

The autoregressive coefficients $a_j$ ($j = 0, \ldots N$) and $b_{j,k}$ ($j, k = 1, \ldots N$) are fit by minimizing the mean-squared difference between the modeled time series $y_n$ and the observed time series $y_n^*$. The minimization process also results in a system of linear equations, which are generalizations of the *Yule-Walker equations* for the linear autoregressive model.

Conceptually, the time series $y_n$ is considered to be the output of a circuit with nonlinear feedback, driven by a noise $x_n$. In principle, the coefficients $b_{j,k}$ describes dynamical features that are not evident in the power spectrum or related measures. Although the equations for the autoregressive coefficients $a_j$ and $b_{j,k}$ are linear, the estimates of these parameters are often unstable, essentially because a large number of them must be estimated often resulting in significant estimation errors. This means that all *linear predictive systems* tend to break down once a rationalization loop has been generated. As parameters of the volatility driving process, which are used to extricate the implied volatility on the longer term options from the implied volatility on the short-term ones, are estimated by an AR1 model, which belongs to the class of regression models collectively referred to as the GLIM (General Linear Model), the parameter estimates go 'out of sync' with those predicted by a theoretical pricing model.

Unfortunately, there is no straightforward method to distinguish linear time series models ($H_0$) from non-linear alternatives ($H_A$). The approach generally taken is to test the $H_0$ of linearity against a pre-chosen particular non-linear $H_A$. Using the classical theory of statistical hypothesis testing, several test statistics have been developed for this purpose. They can be classified as Lagrange Multiplier (LM) tests, likelihood ratio (LR) tests and Wald (W) tests. The LR test requires estimation of the model parameters both under $H_0$ and $H_A$, whereas the LM test requires estimation only under $H_0$. Hence in case of a complicated, non-linear $H_A$ containing many more parameters as compared to the model under $H_0$, the LM test is far more convenient to use. On the other hand, the LM test is designed to reveal specific types of non-linearities. The test may also have some power against inappropriate alternatives. However, there may at the same time exist alternative non-linear models against which an LM test is not powerful. Thus rejecting $H_0$ on the basis of such a test does not permit robust conclusions about the nature of the non-linearity. One possible solution to this problem is using a W test which estimates the model parameters under a well-specified non-linear $H_A$ [14].

## *3. The Zadeh argument revisited*

In the face of non-linear feedback processes generated by *dissonant information sources*, even mathematically sound rule-based reasoning schemes often tend to break down. As a pertinent illustration, we take Zadeh's argument against the well-known Dempster's rule [15] [16]. Let $\Theta = \{\theta_1, \theta_2 \ldots \theta_n\}$ stand for a set of n mutually exhaustive, elementary events that cannot be precisely defined and classified making it impossible to construct a larger set $\Theta_{ref}$ of disjoint elementary hypotheses.

The assumption of exhaustiveness is not a strong one because whenever $\theta_j$, $j = 1, 2 \ldots n$ does not constitute an exhaustive set of elementary events, one can always add an extra element $\theta_0$ such that $\theta_j$, $j = 0, 1 \ldots n$ describes an exhaustive set. Then, if $\Theta$ is considered to be a *general frame of discernment* of the problem under consideration, a map **m (.): $D^\Theta$ → [0, 1]** may be defined associated with a given body of evidence $B$ that can support paradoxical information as follows:

$$m(\phi) = 0 \quad (5)$$
$$\Sigma_{A \in D^\Theta} m(A) = 1 \quad (6)$$

Then m (A) is called A's *basic probability number*. In line with the *Dempster-Shafer Theory*, the *belief* and *plausibility functions* are defined as follows:

$$\mathbf{Bel}(A) = \Sigma_{B \in D^\Theta, B \subseteq A} \, m(B) \quad (7)$$
$$\mathbf{Pl}(A) = \Sigma_{B \in D^\Theta, B \cap A \neq \phi} \, m(B) \quad (8)$$

Now let $Bel_1$ (.) and $Bel_2$ (.) be two belief functions over the same frame of discernment $\Theta$ and their corresponding *information granules* $m_1$ (.) and $m_2$ (.). Then the combined *global belief function* is obtained as **Bel$_1$ (.) = Bel$_1$ (.) ⊕ Bel$_2$ (.)** by combining the information granules $m_1$ (.) and $m_2$ (.) as follows for $m(\phi) = 0$ and for any $C \neq 0$ and $C \subseteq \Theta$;

**[m$_1$ ⊕ m$_2$] (C) = [$\Sigma_{A \cap B = C}$ m$_1$ (A) m$_2$ (B)] / [1 - $\Sigma_{A \cap B = \phi}$ m$_1$ (A) m$_2$ (B)] (9)**

The summation notation $\Sigma_{A \cap B = C}$ is necessarily interpreted as the sum over all $A, B \subseteq \Theta$ such that $A \cap B = C$. The orthogonal sum m (.) is considered a basic probability assignment if and only if the denominator in equation (5) is non-zero. Otherwise the orthogonal sum m (.) does not exist and the bodies of evidences $B_1$ and $B_2$ are said to be in *full contradiction*.

Such a case can arise when there exists A ⊂ Θ such that $Bel_1$ (A) =1 and $Bel_2$ ($A^c$) = 1 – a problem associated with *optimal Bayesian information fusion rule* (Dezert, 2001). Extending Zadeh's argument to option market anomalies, if we now assume that under conditions of *asymmetric market information*, two market players with *homogeneous expectations* view implied volatility on the long-term options. One of them sees it as either arising out of (A) current excursion in implied volatility on short-term options with probability 0.99 or out of (C) random white noise with probability of 0.01. The other sees it as either arising out of (B) historical pattern of implied volatility on short-run options with probability 0.99 or out of (C) random white noise with probability of 0.01.

Using Dempster's rule of combination, the unexpected final conclusion boils down to the expression m (C) = [m1 ⊕ m2] (C) = 0.0001/(1 – 0.0099 – 0.0099 – 0.9801) = 1 i.e. the determinant of implied volatility on long-run options is random white noise with absolute certainty!

To deal with this information fusion problem a new combination rule has been proposed under the name of *Dezert-Smarandache combination rule of paradoxical sources of evidence*, which looks for the optimal combination i.e. the basic probability assignment **m (.) = m1 (.) ⊕ m2 (.)** that *maximizes the joint entropy* of the two information sources [17].

The Zadeh illustration originally sought to bring out the fallacy of automated reasoning based on the Dempster's rule and showed that some form of the *degree of conflict* between the sources must be considered before applying the rule. However, in the context of financial markets this assumes a great amount of practical significance in terms of how it might explain some of the recurrent anomalies in rule-based information processing by inter-related market players in the face of apparently conflicting knowledge sources. The traditional conflict between the *fundamental analysts* and the *technical analysts* over the credibility of their respective knowledge sources is of course all too well known!

*4. Market Information Reconciliation Based on the Concept of* **Neutrosophic Risk**

Neutrosophy is a new branch of philosophy that is concerned with *neutralities* and their interaction with various ideational spectra. Let T, I, F be real subsets of the *non-standard interval* $]^-0, 1^+[$. If $\varepsilon > 0$ is an infinitesimal such that for all positive integers n and we have $|\varepsilon| < 1/n$, then the non-standard finite numbers $1^+ = 1+\varepsilon$ and $0^- = 0-\varepsilon$ form the boundaries of the non-standard interval $]^-0, 1^+[$. Statically, T, I, F are *subsets* while dynamically they may be viewed as *set-valued vector functions*. If a logical proposition is said to be t% true in T, i% indeterminate in I and f% false in F then T, I, F are referred to as the *neutrosophic components*. Neutrosophic probability is useful to events that are shrouded in a *veil of indeterminacy* like the *actual* implied volatility of long-term options. As this approach uses a *subset-approximation* for truth-values, indeterminacy and falsity-values it provides a better approximation than classical probability to uncertain events.

The neutrosophic probability approach also makes a distinction between "relative sure event", event that is true only in certain world(s): NP (rse) = 1, and "absolute sure event", event that is true for all possible world(s): NP (ase) =$1^+$. Similar relations can be drawn for "relative impossible event" / "absolute impossible event" and "relative indeterminate event" / "absolute indeterminate event". In case where the truth- and falsity-components are complimentary i.e. they sum up to unity, and there is no indeterminacy and one is reduced to classical probability. Therefore, neutrosophic probability may be viewed as a generalization of classical and imprecise probabilities. [18]

When a long-term option priced by the collective action of the market players is observed to be deviating from the theoretical price, three possibilities must be considered:

(1) The theoretical price is obtained by an inadequate pricing model, which means that the market price may well *be* the true price,

(2) An unstable rationalization loop has taken shape that has pushed the market price of the option 'out of sync' with its true price, or

(3) The nature of the deviation is indeterminate and could be due to either (a) or (b) or a super-position of both (a) and (b) and/or due to some random white noise.

However, it is to be noted that in none of these three possible cases are we referring to the efficiency or otherwise of the market as a whole. The market can only be as efficient as the information it gets to process. We term the systematic risk associated with the efficient market as *resolvable risk*. Therefore, if the information about the true price of the option is misleading (perhaps due to an inadequate pricing model), the market cannot be expected to process it into something useful – after all, the markets can't be expected to pull jack-rabbits out of empty hats! The perceived risk resulting from the imprecision associated with how human psycho-cognitive factors subjectively interpret information and use the processed information in decision-making is what we term as *irresolvable* (or *neutrosophic*) *risk*.

With T, I, F as the neutrosophic components, let us now define the following events:

**H = {p: p is the true option price determined by the theoretical pricing model}** and

**M = {p: p is the true option price determined by the prevailing market price} (10)**

Then there is a t% chance that the event (H ∩ $M^c$) is true, or corollarily, the corresponding complimentary event ($H^c$ ∩ M) is untrue, there is a f% chance that the event ($M^c$ ∩ H) is untrue, or corollarily, the complimentary event (M ∩ $H^c$) is true and there is a i% chance that neither (H ∩ $M^c$) nor (M ∩ $H^c$) is true/untrue; i.e. the determinant of the true market price is indeterminate. This would fit in nicely with possibility (c) enumerated above – that the nature of the deviation could be due to either (a) or (b) or a super-position of both (a) and (b) and/or due to some random white noise.

Illustratively, a set of AR1 models used to extract the mean reversion parameter driving the volatility process over time have *coefficients of determination* in the range say between 50%-70%, then we can say that t varies in the set T (50% - 70%). If the subjective probability assessments of well-informed market players about the weight of the current excursions in implied volatility on short-term options lie in the range say between 40%-60%, then f varies in the set F (40% - 60%). Then unexplained variation in the temporal volatility driving process together with the subjective assessment by the market players will make the event indeterminate by either 30% or 40%. Then the neutrosophic probability of the true price of the option being determined by the theoretical pricing model is **NP (H ∩ M$^c$) = [(50 – 70), (40 – 60), {30, 40}]**.

## 5. Conclusion

Finally, in terms of our behavioral conceptualization of the market anomaly primarily as manifestation of mass cognitive dissonance, the joint neutrosophic probability NP (H ∩ M$^c$) will also be indicative of the extent to which an unstable rationalization loop has formed out of such mass cognitive dissonance that is causing the market price to deviate from the true price of the option. Obviously increasing strength of the non-linear feedback process fuelling the rationalization loop will tend to increase this deviation. As human psychology; and consequently a lot of subjectivity; is involved in the process of determining what drives the market prices, neutrosophic reasoning will tend to reconcile market information much more realistically than classical probability theory. Neutrosophic reasoning approach will also be an improvement over rule-based reasoning possibly avoiding pitfalls like that brought out by Zadeh's argument. This has particularly significant implications for the vast majority of market players who rely on signals generated by some automated trading system following simple rule-based logic.

However, the fact that there is inherent subjectivity in processing the price information coming out of financial markets, given that the way a particular piece of information is subjectively interpreted by an individual investor may not be the globally correct interpretation, there is always the matter of irresolvable risk that will tend to pre-dispose the investor in favour of some *safe* investment alternative that offers some protection against both resolvable as well as irresolvable risk. This highlights the rapidly increasing importance and popularity of *safe* investment options that are based on some form of *portfolio insurance* i.e. an investment mechanism where the investor has some kind of in-built downside protection against adverse price movements resulting from erroneous interpretation of market information e.g. constant proportion portfolio insurance (CPPI) and its generalized form – options based portfolio insurance (OBPI). Such portfolio insurance strategies offer protection against all possible downsides, whether resulting out of resolvable or irresolvable risk, thereby making the investors feel confident about the decisions they take.

# How Extreme Events Can Affect a Seemingly Stabilized Population: a Stochastic Rendition of Ricker's Model

S. Bhattacharya
Department of Business Administration
Alaska Pacific University, U.S.A.
E-mail: sbhattacharya@alaskapacific.edu

S. Malakar
Department of Chemistry and Biochemistry
University of Alaska, U.S.A.

F. Smarandache
Department of Mathematics
University of New Mexico, U.S.A.

**Abstract**

Our paper computationally explores Ricker's *predator satiation* model with the objective of studying how the extinction dynamics of an animal species having a two-stage life-cycle is affected by a sudden spike in mortality due to an extraneous extreme event. Our simulation model has been designed and implemented using sockeye salmon population data based on a stochastic version of Ricker's model; with the shock size being reflected by a sudden reduction in the carrying capacity of the environment for this species. Our results show that even for a relatively marginal increase in the negative impact of an extreme event on the carrying capacity of the environment, a species with an otherwise stable population may be driven close to extinction.

**Key words:** Ricker's model, extinction dynamics, extreme event, Monte Carlo simulation

**Background and research objective**

PVA approaches do not normally consider the risk of catastrophic extreme events under the pretext that no population size can be large enough to guarantee survival of a species in the event of a large-scale natural catastrophe. [1] Nevertheless, it is only very intuitive that some species are more "delicate" than others; and although not presently under any clearly observed threat, could become threatened with extinction very quickly if an *extreme event* was to occur even on a low-to-moderate scale. The term "extreme event" is preferred to "catastrophe" because catastrophe usually implies a natural event whereas; quite clearly; the chance of man-caused extreme events poses a much greater threat at present to a number of animal species as compared to any large-scale natural catastrophe.

An animal has a two-stage life cycle when; in the first stage, newborns become immature youths and in the second stage; the immature youths become mature adults. Therefore, in terms of the stage-specific approach, if $Y_t$ denotes the number of immature young in stage t and $A_t$ denotes the number of mature adults, then the number of adults in year t + 1 will be some proportion of the young, specifically those that survive to the next (reproductive) stage. Then the formal relationship between the number of mature adults in the next stage and the number of immature youths at present may be written as follows:

$$A_{t+1} = \lambda Y_t$$

Here $\lambda$ is the survival probability, i.e. it is the probability of survival of a youth to maturity. The number of young next year will depend on the number of adults in t:

$$Y_{t+1} = f(A_t)$$

Here f describes the reproduction relation between mature adults and next year's young.

This is a straightforward system of simultaneous difference equations which may be analytically solved using a variation of the *cobwebbing approach*. [2] The solution process begins with an initial point $(Y_1, A_1)$ and iteratively determines the next point $(Y_2, A_2)$. If *predator satiation* is built into the process, then we simply end up with Ricker's model:

$$Y_{t+1} = \alpha A_t e^{-A_t/K}$$

Here $\alpha$ is the maximum reproduction rate (for an initial small population) and K is the population size at which the reproduction rate is approximately half its maximum [3]. Putting $\beta = 1/K$ we can re-write Ricker's equation as follows:

$$Y_{t+1} = \alpha A_t e^{-\beta A_t}$$

It has been shown that if $(Y_0, A_0)$ lies within the first of three possible ranges, $(Y_n, A_n)$ approaches (0, 0) in successive years and the population becomes extinct. If $(Y_0, A_0)$ lies within the third range then $(Y_n, A_n)$ equilibrate to a steady-state value of $(Y^*, A^*)$. Populations that begin with $(Y_0, A_0)$ within the second range oscillate between $(Y^*, 0)$ and $(0, A^*)$. Such alternating behavior indicates one of the year classes, or cohorts, become extinct while the other persists i.e. adult breeding stock appear only every other year. Thus the model reveals that three quite different results occur depending initially only on the starting sizes of the population and its distribution among the two stages. [4]

We use the same basic model in our research but instead of analytically solving the system of difference equations, we use the same to simulate the population dynamics as a stochastic process implemented on an MS-Excel spreadsheet. Rather than using a closed-form equation like Ricker's model to represent the functional relationship between $Y_{t+1}$ and $A_t$, we use a Monte Carlo method to simulate the stage-transition process within Ricker's framework; introducing a massive perturbation with a very small probability in order to emulate a catastrophic event.[5]

**Conceptual framework**

We have a formulated a stochastic population growth model with an inbuilt capacity to generate an extreme event based on a theoretical probability distribution. The non-stochastic part of the model corresponds to Ricker's relationship between $Y_{t+1}$ and $A_t$. The stochastic part has to do with whether or not an extreme event occurs at a particular time point. The gamma distribution has been chosen to make the probability distribution for the extreme event a skewed one as it is likely to be in reality. Instead of analytically solving the system of simultaneous difference equations iteratively in some variation of the cobwebbing method, we have used them in a spreadsheet model to simulate the population growth over a span of ten time periods.

We apply a computational methodology whereby the initial number of immature young is hypothesized to either attain the expected number predicted by Ricker's model or drastically fall below that number at the end of every stage, depending on whether an extraneous extreme event does not occur or actually occurs. The mortalities as a result of an extreme event at any time point is expressed as a percentage of the pristine population size for a clearer comparative view.

**Model building**

Among various faunal species, the population dynamics of the sockeye salmon (*oncorhynchus nerka*) has been most extensively studied using Rickert's model. Salmon are unique in that they breed in particular fresh water systems before they die. Their offspring migrates to the ocean and upon reproductive maturity, they are guided by a hitherto unaccounted instinctive drive to swim back to the very same fresh waters where they were born to spawn their own offspring and perish. Salmon populations thus are very sensitive to habitat changes and human activities that have a negative impact on riparian ecosystems that serve as breeding grounds for salmon can adversely affect the peculiar life-cycle of the salmon. Many of the ancient salmon runs (notably those in California river systems) have now gone extinct and it is our hypothesis that an even seemingly stabilized population can be rapidly driven to extinction due to the effect of an extraneous (quite possibly man-made) extreme event with the capacity to cause mass mortality. The following table shows the four-year averages of the sockeye salmon population in the Skeena river system in British Columbia in the first half of the twentieth century.

| Year | Population (in thousands) |
|------|---------------------------|
| 1908 | 1,098 |
| 1912 | 740 |
| 1916 | 714 |
| 1920 | 615 |
| 1924 | 706 |
| 1928 | 510 |
| 1932 | 278 |
| 1936 | 448 |
| 1940 | 528 |
| 1944 | 639 |
| 1948 | 523 |

(*Source*: http://www-rohan.sdsu.edu/~jmahaffy/courses/s00/math121/lectures/product_rule/product.html#Ricker'sModel)

A non-linear least squares best-fit to Ricker's model is obtained for the above set of data is obtained as follows:

$$\text{Minimize } \varepsilon^2 = \sum_{t=1}^{n}[d_t - \{\alpha A_t e^{-\beta A_t}\}]^2, \text{ where } d_t \text{ is the actual population size in year t.}$$

The necessary conditions to the above least squares best-fit problem is obtained as follows:

$$\partial(\varepsilon^2)/\partial\alpha = \partial(\varepsilon^2)/\partial\beta = 0; \text{ whereby we get } \alpha^* \approx 1.54 \text{ and } \beta^* \approx 7.8 \times 10^{-4}$$

Plugging these parameters into Ricker's model indeed yields a fairly good approximation of the salmon population stabilization in the Skeena river system in the first half of the previous century.

As the probability distribution of an extraneous extreme event is likely to be a highly skewed one, we have generated our random variables from the cumulative distribution function (*cdf*) of the gamma distribution rather than the normal distribution. The distribution boundaries are fixed by generating random integers in the range 1 to 100 and using these random integers to define the shape and scale parameters of the gamma distribution. The gamma distribution performs better than the normal distribution when the distribution to be matched is highly right-skewed; as is desired in our model. The combination of a large variance and a lower limit at zero makes fitting a normal distribution rather unsuitable in such cases.[6] The probability density function of the gamma distribution is given as follows:

$$f(x, a, b) = \{b^a \Gamma(a)\}^{-1} x^{a-1} e^{-x/b} \text{ for } x > 0$$

Here α > 0 is the shape parameter and β > 0 is the scale parameter of the gamma distribution. The cumulative distribution function may be expressed in terms of the *incomplete gamma function* as follows:

$$F(x, a, b) = \int_0^x f(u)du = \gamma(a, x/b)/\Gamma(a)$$

In our spreadsheet model, we have $F(R, R/2, 2)$ as our *cdf* of the gamma distribution. Here $R$ is an integer randomly sampled from the range 1 to 100. An interesting statistical result of having these values for x, α and β is that the cumulative gamma distribution value becomes equalized with the value $[1 - \chi^2(R)]$ having $R$ degrees of freedom, thus allowing $\chi^2$ *goodness-of-fit* tests. [7]

Our model is specifically designed to simulate the extinction dynamics of sockeye salmon population using a stochastic version of Ricker's model; with the shock size being based on a sudden reduction in the parameter K i.e. the carrying capacity of the environment for this species. The model parameters are same as those of Ricker's model i.e. α and β (which is the reciprocal of K). We have kept α constant at all times at 1.54, which was the least squares best-fit value obtained for that parameter. We have kept a β of 0.00078 (i.e. the best-fit value) when no extreme event occurs and have varied the β between 0.00195 and 0.0156 (i.e. between 2.5 times to 20 times the best-fit value) for cases where an extreme event occurred. We have a third parameter $c$ which is basically a 'switching constant' that determines whether an extreme event occurs or not. The switch is turned on triggering an extreme event when a random draw from a cumulative gamma distribution yields a value less than or equal to $c$. Using $F(R, R/2, 2)$ as our *cdf* of the gamma distribution where $R$ is a randomly drawn integer in the range (1, 100) means that the cumulative gamma function will randomly select from the approximate interval 0.518 ~ 0.683. By fixing the value of $c$ at 0.5189 in our model we have effectively reduced the probability of occurrence of an extreme event to a miniscule magnitude relative to that of an extreme event not occurring. We have used the sockeye salmon population data from the table presented earlier For each level of the β parameter, we simulated the system and observed the maximum possible number of mortalities from an extreme event at that level of β. The results are reported below.

**Results obtained from the simulation model**

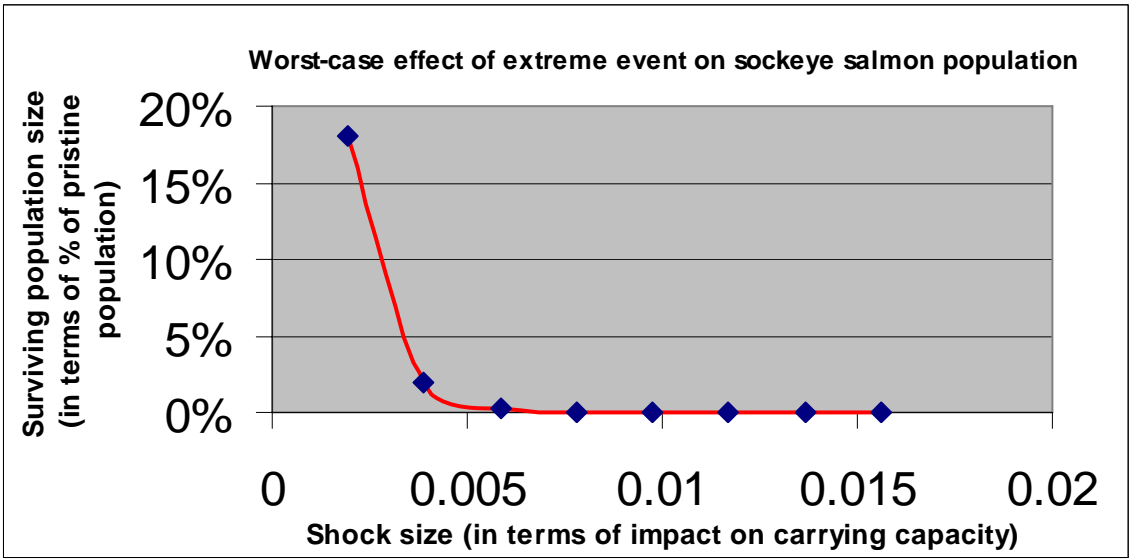

We made 100 independent simulation runs for each of the eight levels of β. The low probability of extreme event assigned in our study yielded a mean of 1.375 for the number of observed worst-case scenarios (i.e. situations of maximum mortality) with a standard deviation of approximately 0.92. The worst-case scenarios for our choice of parameters necessarily occur if the extreme event occurs in the first time point when the species population is at its maximum size. Our model shows that in worst-case scenarios, the size of surviving population after an extreme event that could seed the ultimate recovery of the species to pre-catastrophe numbers (staying within the broad framework of Ricker's model) drops from about 18% of the pristine population size for a shock size corresponding to 2.5 times the best-fit β; to only about 0.000005% of the pristine population size for a shock size corresponding to 20 times the best-fit β.

Therefore, if the minimum required size of the surviving population is at least say 20% of the pristine population in order to survive and recover to pre-catastrophe numbers, the species could go extinct if an extreme event caused a little more than two-fold decrease in the environmental carrying capacity! Even if the minimum required size for recovery was relatively low at say around 2% of the pristine population, an extreme event that caused a five-fold decrease in the environmental carrying capacity could very easily force the species to the brink of extinction. An immediate course of future extension of our work would be allowing the fecundity parameter α to be affected by extreme events as is very likely in case of say a large-scale chemical contamination of an ecosystem due to a faulty industrial waste-treatment facility.

**Conclusion**

Our study has shown that even for a relatively marginal 2.5-fold decrease in the environmental carrying capacity due to an extreme event, a worst-case scenario could mean a mortality figure well above 80% of the pristine population. As a guide for future PVA studies we may suggest that one should not be deterred simply by the notion that extreme events are uncontrollable and hence outside the purview of computational modeling. Indeed the effect of an extreme event can almost always prove to be fatal for a species but nevertheless, as our study shows, there is ample scope and justification for future scientific enquiries into the relationship between survival probability of a species and the adverse impact of an extreme event on ecological sustainability.

# Processing Uncertainty and Indeterminacy in Information Systems Projects Success Mapping

Jose L. Salmeron

Pablo de Olavide University at Seville

Spain

Florentin Smarandache

University of New Mexico

Gallup, USA

**Abstract**

IS projects success is a complex concept, and its evaluation is complicated, unstructured and not readily quantifiable. Numerous scientific publications address the issue of success in the IS field as well as in other fields. But, little efforts have been done for processing indeterminacy and uncertainty in success research. This paper shows a formal method for mapping success using Neutrosophic Success Map. This is an emerging tool for processing indeterminacy and uncertainty in success research. EIS success have been analyzed using this tool.

**Keywords**: Indeterminacy, Uncertainty, Information Systems Success, Neutrosophic logic, Neutrosophic Cognitive Maps, Fuzzy logic, Fuzzy Cognitive Maps.

## 1. Introduction

For academics and practitioners concerned with computer-based Information Systems (IS), one central issue is the study of development and implementation project success. Literature (Barros et al., 2004; Poon and Wagner, 2001; Rainer and Watson, 1995;

Redmil, 1990) suggest that IS projects have lower success rates than other technical projects. Irrespective of the accuracy of this presumption, the number of unsuccessful IS projects are over the number of successful ones. Therefore, it is worthwhile to develop a formal method for mapping success, since proper comprehension of the complex nature of IS success is critical for the successful application of technical principles to this discipline.

To increase the chances of an IS project to be perceived as successful for people involved in project, it is necessary to identify at the outset of the project what factors are important and influencing that success. These are the Critical Success Factors (CSF) of the project. Whereas several CSF analyses appear in the literature, most of them do not have any technical background. In addition, almost none of them focus on relations between them. In addition, it is important to discover the relationships between them. Research about it was becoming scarce.

In this paper, we propose the use of an innovative technique for processing uncertainty and indeterminacy to set success maps in IS projects. The main strengths of this paper are two-folds: it provides a method for processing indeterminacy and uncertainty within success and it also allows building a success map.

The remainder of this paper is structured as follows: Section 2 shows previous research; Section 3 reviews cognitive maps and its evolution; Section 4 is focused on the research model; Section 5 presents and analyzes the results; the final section shows the paper's conclusions.

## 2. Previous research

Success is not depending to just one issue. Complex relations of interdependence exist between IS, organization, and users. Thus, for example, reducing costs in an organization cannot be derived solely from IS implementation. Studies indicate that the IS success is hard to assess because it represent a vague topic that does not easily lend itself to direct measurement (DeLone and McLean, 1992).

According to Zviran and Erlich (2003), academics tried to assess the IS success as a function of cost-benefit (King and Schrems, 1978), information value (Epstein and King, 1983; Gallagher, 1974), or organization performance (Turner, 1982). System acceptance (Davis, 1989) has used for it too. Anyway, cost-benefit, information value, system acceptance, and organization performance are difficult to apply as measures.

Critical Success Factor method (Rockart, 1979) have been used as a mean for identifying the important elements of IS success since 1979. It was developed as a method to enable CEOs to recognize their own information needs so that IS could be built to meet those needs. This concept has received a wide diffusion among IS scholars and practitioners (Butler and Fitzgerald, 1999).

Numerous scientific publications address the issue of CSF in the IS field as well as in other fields. But, little efforts have been done for introducing formal methods in success research. Some authors (Poon and Wagner, 2001) analysed some aspects of CSF just by the use of personal interviews, whereas others (Ragahunathan et al., 1989) carried out a Survey-based field study. Interviews and/or questionnaires are common tools for measuring success. However, formal methodology is not usual.

On the other hand, Salmeron and Herrero (2005) propose a hierarchical model to model success. Anyway, indeterminacy was not processed. Therefore, we think that a formal method to process indeterminacy and uncertainty in IS success is an useful endeavour.

## 3. Uncertainty and Indeterminacy processing in cognitive maps

### 3.1. Cognitive mapping

A cognitive map shows a representation of how humans think about a particular issue, by analyzing, arranging the problems and graphically mapping concepts that are connected between them. In addition, it identifies causes and effects and explains causal links (Eden and Ackermann, 1992). The cognitive maps study perceptions about the world and the way they act to reach human desires within their world. Kelly (1955, 1970) gives the foundation for this theory, based on a particular cognitive psychological body of knowledge. The base postulate for the theory is that "a person's processes are psychologically canalized by the ways in which he anticipates events." Mental models of top managers in firms operating in a competitive environment have been studied (Barr et al., 1992) using cognitive mapping. They suggest that the cognitive models of these managers must take into account significant new areas of opportunity or technological developments, if they want stay ahead. In this sense, it is critical to consider mental models in success research.

### 3.2. Neutrosophic Cognitive Maps (NCM)

In fact, success is a complex concept, and its evaluation is complicated, unstructured and not readily quantifiable. The NCM model seems to be a good choice to deal with this ambiguity. NCM are flexible and can be customised in order to consider the CSFs of different IT projects.

Neutrosophic Cognitive Maps (Vasantha-Kandasamy and Smarandache, 2003) is based on Neutrosophic Logic (Smarandache, 1999) and Fuzzy Cognitive Maps. Neutrosophic Logic emerges as an alternative to the existing logics and it represents a mathematical model of uncertainty, and indeterminacy. A logic in which each proposition is estimated to have the percentage of truth in a subset T, the percentage of indeterminacy in a subset I, and the percentage of falsity in a subset F, is called Neutrosophic Logic. It uses a subset of truth (or indeterminacy, or falsity), instead of using a number, because in many cases, humans are not able to exactly determine the percentages of truth and of falsity but to approximate them: for example a proposition is between 30-40% true. The subsets are not necessarily intervals, but any sets (discrete, continuous, open or closed or half-open/ half-closed interval, intersections or unions of the previous sets, etc.) in accordance with the given proposition. A subset may have one element only in special cases of this logic. It is imperative to mention here that the Neutrosophic logic is a strait generalization of the theory of Intuitionistic Fuzzy Logic.

Neutrosophic Logic which is an extension/combination of the fuzzy logic in which indeterminacy is included. It has become very essential that the notion of neutrosophic logic play a vital role in several of the real world problems like law, medicine, industry, finance, IT, stocks and share, and so on. Fuzzy theory measures the grade of membership or the non-existence of a membership in the revolutionary way but fuzzy theory has failed to attribute the concept when the relations between notions or nodes or concepts in problems are indeterminate. In fact one can say the inclusion of the concept of indeterminate situation with fuzzy concepts will form the neutrosophic concepts (there also is the neutrosophic set, neutrosophic probability and statistics).

In this sense, Fuzzy Cognitive Maps mainly deal with the relation / non-relation between two nodes or concepts but it fails to deal with the relation between two conceptual nodes when the relation is an indeterminate one. Neutrosophic logic is the only tool known to us, which deals with the notions of indeterminacy.

A Neutrosophic Cognitive Map (NCM) is a neutrosophic directed graph with concepts like policies, events, etc. as nodes and causalities or indeterminates as edges. It represents the causal relationship between concepts. A neutrosophic directed graph is a directed graph in which at least one edge is an indeterminacy denoted by dotted lines.

Let $C_1, C_2,\ldots, C_n$ denote n nodes, further we assume each node is a neutrosophic vector from neutrosophic vector space V. So a node $C_i$ will be represented by $(x_1, \ldots, x_n)$ where $x_k$'s are zero or one or I (I is the indeterminate introduced before) and $x_k = 1$ means that the node $C_k$ is in the on state and $x_k = 0$ means the node is in the off state and $x_k = I$ means the nodes state is an indeterminate at that time or in that situation.

Let $C_i$ and $C_j$ denote the two nodes of the NCM. The directed edge from $C_i$ to $C_j$ denotes the causality of $C_i$ on $C_j$ called connections. Every edge in the NCM is weighted with a number in the set {-1, 0, 1, I}. Let $e_{ij}$ be the weight of the directed edge $C_iC_j$, $e_{ij} \in \{-1,0,1,I\}$. $e_{ij} = 0$ if $C_i$ does not have any effect on $C_j$, $e_{ij} = 1$ if increase (or decrease) in $C_i$ causes increase (or decreases) in $C_j$, $e_{ij} = -1$ if increase (or decrease) in $C_i$ causes decrease (or increase) in $C_j$. $e_{ij} = I$ if the relation or effect of $C_i$ on $C_j$ is an indeterminate.

The edge $e_{ij}$ takes values in the fuzzy causal interval $[-1, 1]$ ($e_{ij} = 0$ indicates no causality, $e_{ij} > 0$ indicates causal increase; that $C_j$ increases as $C_i$ increases and $C_j$ decreases as $C_i$ decreases, $e_{ij} < 0$ indicates causal decrease or negative causality $C_j$ decreases as $C_i$ increases or $C_j$, increases as $C_i$ decreases. Simple FCMs have edge value in {-1, 0, 1}. Thus if causality occurs it occurs to maximal positive or negative degree.

It is important to note that $e_{ij}$ measures only absence or presence of influence of the node $C_i$ on $C_j$ but till now any researcher has not contemplated the indeterminacy of any relation between two nodes $C_i$ and $C_j$. When we deal with unsupervised data, there are situations when no relation can be determined between some two nodes. In our view this will certainly give a more appropriate result and also caution the user about the risks and opportunities of indeterminacy.

Using NCM is possible to build a Neutrosophic Success Map (NSM). NSM nodes represent Critical Success Factors (CSF). They are the limited number of areas in which results, if they are satisfactory, will ensure successful competitive performance for the organization. They are the few key areas where "things must go right" for the project (Rockart, 1979). This tool shows the relations and the fuzzy values within in an easy understanding way. This is an useful approach for non-technical decision makers. At the same time, it allows computation as FCM. Figure 1 shows the NSM static context.

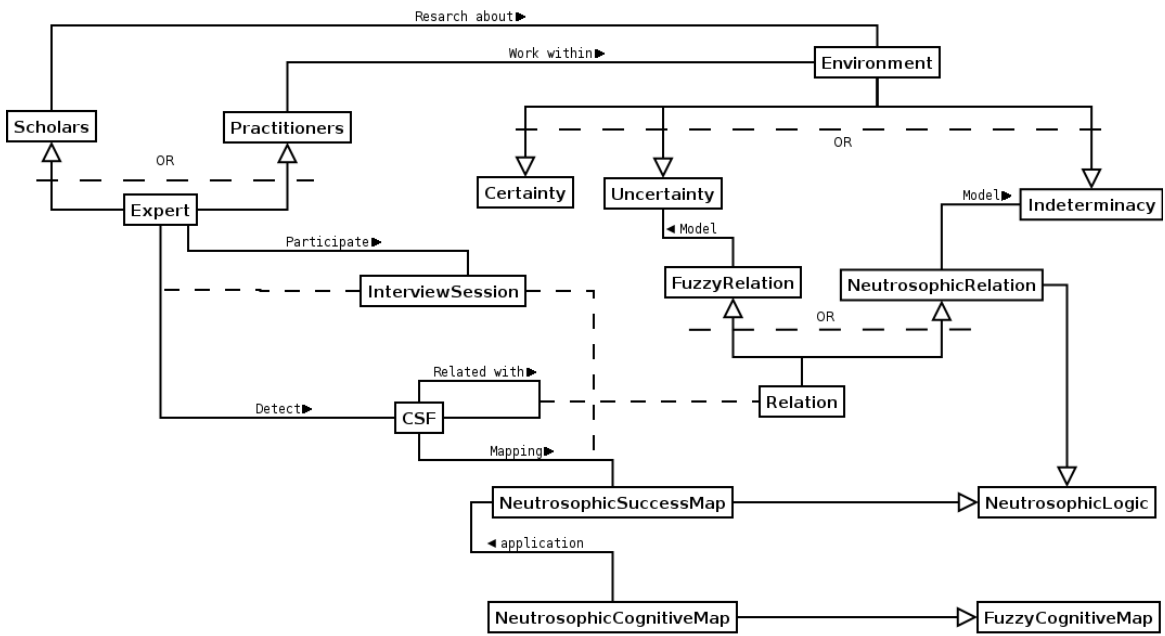

**Figure 1: NSM static context**

**5. Building a NSM**

EIS project have been used for building a Neutrosophic Success Map. EIS, or executive support systems as they are sometimes called, can be defined as computer-based information systems that support communications, coordination, planning and control functions of managers and executives in organizations (Salmeron and Herrero, 2005).

NSM will be based on textual descriptions given by EIS experts on interviews with them. The steps followed are:

1) Experts selection. It is critical step. Expert selection was based on specific knowledge of EIS systems. Experts are 19 EIS users of leading companies and EIS researchers. The composition of the respondents is important. Multiple choices were contemplated. The main selection criterion considered was recognized knowledge in research topic, absence of conflicts of interest and geographic diversity. All conditions were respected. In addition, respondents were not chosen just because they are easily accessible.
2) Identification of CSF influencing the EIS systems.
3) Identification and assess of causal relationships among these CSF. Indeterminacy relations are included.

Experts discover the CSFs and give qualitative estimates of the strengths associated with causal links between nodes representing these CSFs. These estimates, often expressed in imprecise or fuzzy/neutrosophic linguistic terms, are translated into numeric values in the range −1 to 1. In addition, indeterminacy is used for modelling that kind of relations relationships among nodes.

The nodes (CSFs) discover was the following:

1. Users' involvement ($x_1$). It is defined as a mental or psychological state of users toward the system and its development process. It is generally accepted that IS users' involvement in the application design and implementation is important and necessary (Hwang and Thorn, 1999). It is essential in maintenance phase too.
2. Speedy prototype development ($x_2$). It encourages the right information needs because it interacts between user and system as soon as possible.
3. Top management support ($x_3$). EIS support with his/her authority and influence over the rest of the executives.
4. Flexible system ($x_4$). EIS must be flexible enough to be able to get adapted to changes in the types of problems and the needs of information.
5. Right information requirements ($x_5$). Eliciting requirements is one of the most complicated tasks in developing EIS and getting a correct requirement set is challenging.

6. Technological integration ($x_6$). EIS tool selected must be integrated in companies' technological environment.
7. Balanced development team ($x_7$). Suitable human resources are required for developing EIS. Technical background and business knowledge are needed.
8. Business value ($x_8$). The system must solve a critical business problem. There should be a clear business value in EIS use.
9. Change management ($x_9$). It is the process of developing a planned approach to change in a firm. EIS will be a new way of working. Typically the objective is to maximize the collective efforts of all people involved in the change and minimize the risk of failure of EIS project.

The NSM find out is presented in Figure 2. Fuzzy values are included.

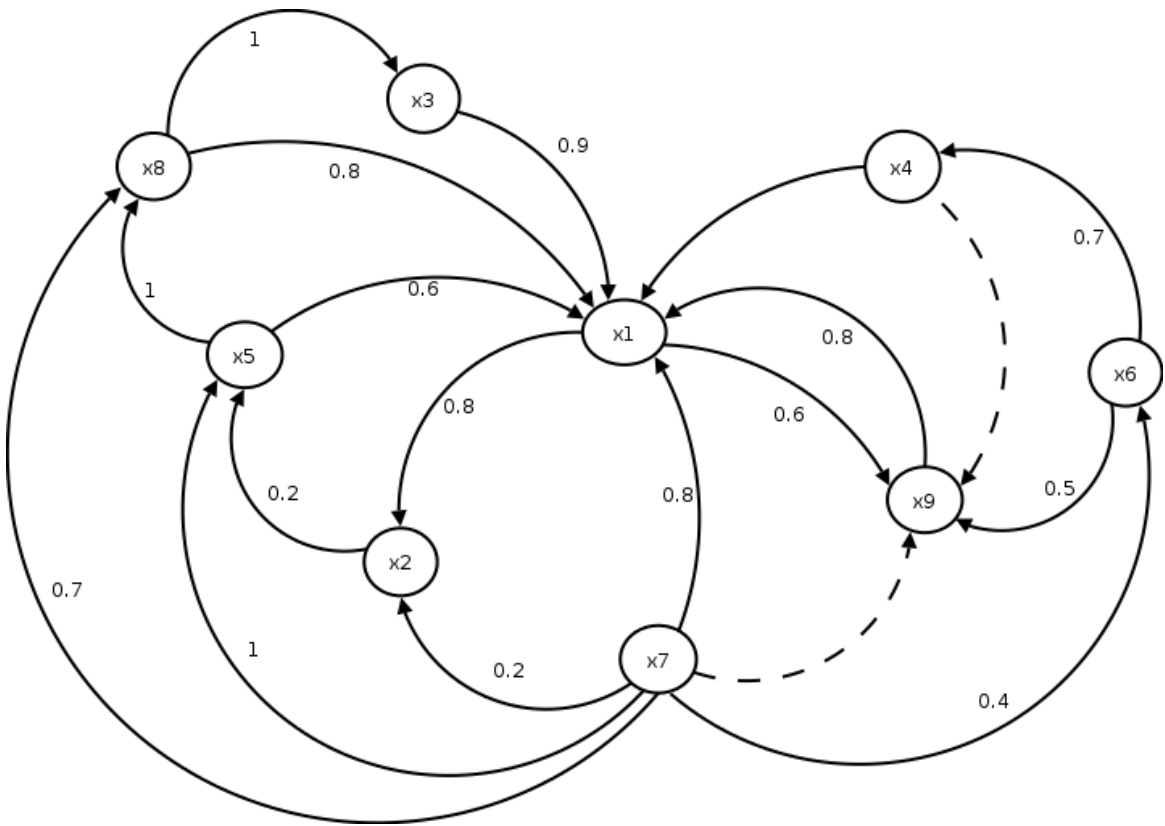

**Figure 2. EIS NSM**

**The adjacency matrix associated to NSM is N(E).**

$$N(E) = \begin{pmatrix} 0 & 0.8 & 0 & 0 & 0 & 0 & 0 & 0 & 0.6 \\ 0 & 0 & 0 & 0 & 0.2 & 0 & 0 & 0 & 0 \\ 0.9 & 0 & 0 & 0 & 0 & 0 & 0 & 0 & 0 \\ 0.1 & 0 & 0 & 0 & 0 & 0 & 0 & 0 & I \\ 0.6 & 0 & 0 & 0 & 0 & 0 & 0 & 1 & 0 \\ 0 & 0 & 0 & 0.7 & 0 & 0 & 0 & 0 & 0.5 \\ 0.8 & 0.2 & 0 & 0 & 1 & 0.4 & 0 & 0.7 & I \\ 0.8 & 0 & 1 & 0 & 0 & 0 & 0 & 0 & 0 \\ 0.8 & 0 & 0 & 0 & 0 & 0 & 0 & 0 & 0 \end{pmatrix}$$

The stronger relations are between $x_5$ to $x_8$, $x_7$ to $x_5$ and $x_8$ to $x_3$. It follows that a balanced development team has a positive influence over elicitation requirements process. In the same sense, eliciting right requirements have a positive influence over system business value and system business value over top management support. In addition, users' involvement receives influence from six nodes.

On the other hand, we have found two neutrosophic relations between $x_4$ to $x_9$, $x_5$ to $x_3$ and $x_7$ to $x_9$. It follows that experts perceive indeterminacy in relations between EIS flexibility and balanced team skills to change management. They can not to assess the relation between them, but they perceive that relation could exist. It is an useful information since the decision-makers can be advised from it. They will be able to be careful with those relations.

In addition, NSM predict effects of one or more CSFs (nodes) in the regarding ones. If we know that any CSFs are on, we can discover the influence over the others. This process is similar in Fuzzy Cognitive Maps.

Let $\overrightarrow{C_1C_2}, \overrightarrow{C_2C_3}, ..., \overrightarrow{C_{n-1}C_n}$ be cycle (Vasantha-Kandasamy and Smarandache, 2003), when $C_i$ is switched on and if the causality flow through the edges of a cycle and if it again causes $C_i$, we say that the dynamical system goes round and round. This is true for any node $C_i$, for i = 1, 2,..., n. the equilibrium state for this dynamical system is called the hidden pattern. If the equilibrium state of a dynamical system is a unique state vector, then it is called a fixed point. If the NSM settles with a state vector repeating in the form

$x_1 \rightarrow x_2 \rightarrow ... \rightarrow x_i \rightarrow x_1$,

then this equilibrium is called a limit cycle of the NSM.

Let $C_1, C_2, \ldots, C_n$ be the CSFs of an NSM. Let E be the associated adjacency matrix. Let us find the hidden pattern when $x_1$ is switched on when an input is given as the vector $A_1 = (1, 0, 0, \ldots, 0)$, the data should pass through the neutrosophic matrix N(E), this is done by multiplying $A_1$ by the matrix N(E). Let $A_1 N(E) = (a_1, a_2, \ldots, a_n)$ with the threshold operation that is by replacing $a_i$ by 1 if $a_i > k$ and $a_i$ by 0 if $a_i < k$ and $a_i$ by I if $a_i$ is not a integer.

$$f(k) = \begin{cases} a_i < k \Rightarrow a_i = 0 \\ a_i > k \Rightarrow a_i = 1 \\ a_i = b + c \times I \Rightarrow a_i = b \\ a_i = c \times I \Rightarrow a_i = I \end{cases}$$

This procedure is repeated till we get a limit cycle or a fixed point. According to this, the limit cycle or a fixed point of vector state of each CSFs is calculated with k=0.5. We take the state vector $A_1 = (1\ 0\ 0\ 0\ 0\ 0\ 0)$. We will see the effect of $A_1$ over the model.

$A_1 N(E) = (0\ \ 0.8\ \ 0\ \ 0\ \ 0\ \ 0\ \ 0\ \ 0\ \ 0.6) \longrightarrow (1\ \ 1\ \ 0\ \ 0\ \ 0\ \ 0\ \ 0\ \ 0\ \ 1) = A_2$

$A_2 N(E) = (0.8\ \ 0.8\ \ 0\ \ 0\ \ 0.2\ \ 0\ \ 0\ \ 0\ \ 0.6) \longrightarrow (1\ \ 1\ \ 0\ \ 0\ \ 0\ \ 0\ \ 0\ \ 0\ \ 1) = A_3 = A_2$

$A_2$ is a fixed point. According with experts the on state of users' involvement has effect over speedy prototype development and change management.

We take the new state vector $A_1 = (1\ 0\ 1\ 0\ 0\ 0\ 0\ 0\ 0)$. We will see the effect of users' involvement and top management support ($A_1$) over the model.

$A_1 N(E) = (0.9\ \ 0.8\ \ 0\ \ 0\ \ 0.2\ \ 0\ \ 0\ \ 0\ \ 0.6) \longrightarrow (1\ \ 1\ \ 1\ \ 0\ \ 0\ \ 0\ \ 0\ \ 0\ \ 1) = A_2$

$A_2 N(E) = (1.7\ \ 0.8\ \ 0\ \ 0\ \ 0.2\ \ 0\ \ 0\ \ 0\ \ 0.6) \longrightarrow (1\ \ 1\ \ 1\ \ 0\ \ 0\ \ 0\ \ 0\ \ 0\ \ 1) = A_3 = A_2$

Thus $A_2=(1\ 1\ 1\ 0\ 0\ 0\ 0\ 0\ 1)$, according with experts the on state of users' involvement and top management support have effects over the prototype speed of development ($x_2$) and change management ($x_9$). It is interesting to discover that both previous state vectors have the same influence over the model. Both vector states have influence over prototype speed of development ($x_2$) and change management, but no direct effect over the rest of CSFs.

The vector states described are only two of the several available, even vectors with several CSFs on. However the proposal here presented is as simple as possible while being consistent with the process, data gathered from the expert's perceptions, and the aims and objectives of the paper.

## 6. Conclusions

The main strengths of this paper are two-folds: it provides a method for project success mapping and it also allows know CSF effects over the other ones. In this paper, we proposed the use of the Neutrosophic Success Maps to map EIS success.

A tool for evaluating suitable success models for IS projects is required due to the increased complexity and uncertainty associated to this kind of projects. This leads to the innovative idea of adapting and improving the existent Neutrosophic theories for their application to indicators of success for IS projects.

Neutrosophic Success Map is an innovative success research approach. NSM is based on Neutrosophic Cognitive Map. The concept of NCM can be used in modelling of systems success, since the concept of indeterminacy play a role in that topic. This was our main aim is to use NCMs in place of FCMs. When an indeterminate causality is present in an FCM we term it as an NCM.

The results not mean that any CSF is unimportant or has not effect over the model. It means what are the respondents' perceptions about the relationships of them. This is a main issue, since it is possible to manage the development process with more information about the expectations of final users.

Anyway, more research is needed about Neutrosophic logic limit and applications. Incorporating the analysis of NCM and NSM, the study proposes an innovative way for success research. We think this is an useful endeavour.

**Computational models pervade all branches of the exact sciences and have in recent times also started to prove to be of immense utility in some of the traditionally 'soft' sciences like ecology, sociology and politics. This volume is a collection of a few cutting-edge research papers on the application of variety of computational models and tools in the analysis, interpretation and solution of vexing real-world problems and issues in economics, management, ecology and global politics by some prolific researchers in the concerned fields.**

**The Editors**